%% file: main.tex
\documentclass[10pt,conference]{IEEEtran}

\def\BibTeX{{\rm B\kern-.05em{\sc i\kern-.025em b}\kern-.08em
    T\kern-.1667em\lower.7ex\hbox{E}\kern-.125emX}}

\input{packages.tex}
\input{acronyms.tex}

\input{commands.tex}

\newcommand{\ra}[1]{\renewcommand{\arraystretch}{#1}}
\definecolor{lightGreen}{HTML}{50847C}
\makeatletter
\newcommand{\linebreakand}{%
  \end{@IEEEauthorhalign}
  \hfill\mbox{}\par
  \mbox{}\hfill\begin{@IEEEauthorhalign}
}
\makeatother

\newtcolorbox{Summary}{
    sharpish corners, 
    boxrule = 0pt, 
    toprule = 3.5pt, 
    toptitle = 1mm, 
    enhanced,
    fuzzy shadow = {0pt}{-2pt}{-0.5pt}{0.5pt}{black!35}, 
    colback = white, 
    colframe = gray, 
    coltitle=black, 
    fonttitle=\bfseries 
}

\begin{document}

\title{LLMs in the Heart of Differential Testing:\\A Case Study on a Medical Rule Engine}



\author{\IEEEauthorblockN{Erblin Isaku}
\IEEEauthorblockA{\textit{Simula Research Laboratory and } \\
\textit{University of Oslo}\\
Oslo, Norway\\
erblin@simula.no}
\and
\IEEEauthorblockN{Christoph Laaber}
\IEEEauthorblockA{\textit{Simula Research Laboratory} \\
Oslo, Norway \\
laaber@simula.no}
\and
\IEEEauthorblockN{Hassan Sartaj}
\IEEEauthorblockA{\textit{Simula Research Laboratory} \\
Oslo, Norway \\
hassan@simula.no}
\linebreakand 
\IEEEauthorblockN{Shaukat Ali}
\IEEEauthorblockA{\textit{Simula Research Laboratory } \\
Oslo, Norway \\
shaukat@simula.no}
\and
\centering
\IEEEauthorblockN{Thomas Schwitalla}
\IEEEauthorblockA{\textit{Cancer Registry of Norway} \\
Oslo, Norway \\
thsc@kreftregisteret.no}
\and
\IEEEauthorblockN{Jan F. Nygård}
\IEEEauthorblockA{\textit{Cancer Registry of Norway and } \\
\textit{UiT The Arctic University of Norway}\\
Oslo, Norway \\
jfn@kreftregisteret.no}
}

\maketitle

\begin{abstract}

The \gls{crn} uses an automated \gls{caress} to support core cancer registry activities, i.e., data capture, data curation, and producing data products and statistics for various stakeholders. 
\guri{} is a core component of \gls{caress}, which is responsible for validating incoming data with medical rules.
Such medical rules are manually implemented by medical experts based on medical standards, regulations, and research.
Since \glspl{llm} have been trained on a large amount of public information, including these documents, they can be employed to generate tests for \guri{}.
Thus, we propose an \gls{llm}-based test generation and differential testing approach (\ourApproach{}) to test \guri{}. 
We experimented with \numllms{} different \glspl{llm}, two medical rule engine implementations, and \originalRules{} real medical rules to investigate the hallucination, success, time efficiency, and robustness of the \glspl{llm} to generate tests, and these tests' ability to find potential issues in \guri{}.
Our results showed that \gpt{} hallucinates the least, is the most successful, and is generally the most robust; however, it has the worst time efficiency. 
Our differential testing revealed 22 medical rules where implementation inconsistencies were discovered (e.g., regarding handling rule versions). Finally, we provide insights for practitioners and researchers based on the results.  


\end{abstract}

\begin{IEEEkeywords}
Large Language Models, Medical Rules, Automated Software Testing, Test Generation, Differential Testing, Electronic Health Records
\end{IEEEkeywords}

\glsresetall{}
\section{Introduction}

The \gls{caress} is a real-world sociotechnical software system developed and operated by the \gls{crn}, a public organization in Norway under the \gls{niph}.
\Gls{caress} receives cancer patient data through cancer messages from, e.g., hospitals, laboratories, and other health registries.
It then processes the incoming data to produce specialized data and statistics for its stakeholders, such as the public, researchers, and policymakers.

One core functionality of \gls{caress} is to validate incoming cancer messages and medical data with a dedicated rule validation software system called \guri{}, which checks the correctness of received cancer messages, validates the data with medical rules, aggregates individual cancer messages into cancer cases, and validates the aggregation.
The rules are manually defined by medical coders based on standards and regulations, such as \gls{icd}-10 (10\textsuperscript{th} edition), \gls{icd}-O-2 (\gls{icd} for Oncology, 2\textsuperscript{nd} edition) for solid tumors, and \gls{icd}-O-3 (\gls{icd} for Oncology, 3\textsuperscript{rd} edition) for non-solid tumors.
Extensively testing \guri{} is important since an incorrect implementation would result in erroneous data or statistics, which are used in medical research as well as monitoring and evaluation of cancer care. 

As \glspl{llm} are trained on a vast amount of public information, including the standards and regulations used for specifying \gls{crn}'s medical rules, we employ \glspl{llm} to generate tests from these rules in a differential testing approach called \ourApproach{} to find potential faults in \guri{}.
We assume that utilizing \glspl{llm} is advantageous over other methods (e.g., search or constraint solvers) due to their semantic knowledge of the medical context, ability to generate realistic and diverse inputs, and capability of understanding rule variable names in different languages (i.e., Norwegian in our case).
\ourApproach{} uses an \gls{llm} to generate tests from medical rules (that yield \rulePass{}, \ruleFail{}, and \ruleNotApplied{} results for each rule), which are then executed on \guri{} and a simple reference implementation (\dvarepp{}).
A mismatch between the outputs potentially stems from an implementation issue in \guri{}.

Previously, we studied a state-of-the-art test generation tool (i.e., EvoMaster~\citep{arcuri:19}) applied to \guri{}~\citep{laaber:23a}, reduced its number of executed tests with \gls{ml} classifiers~\citep{isaku:23}, and built cyber-cyber digital twins of \guri{}~\citep{chengjie:23}.
In this paper, we
\begin{inparaenum}
    \item leverage \glspl{llm} to generate valid medical data as test inputs,
    \item address the oracle problem of rule results through differential testing with a reference implementation,
    and
    \item reduce the number of executed tests against \guri{} by only generating three tests per rule.
\end{inparaenum}

We experimented with \ourApproach{} by selecting \numllms{} \glspl{llm} (i.e., \mistral{}, \llama{}, \mixtral{}, and \gpt{}), \originalRules{} real medical rules and their \mutatedRules{} mutations, and two rule engine implementations (i.e., \guri{} and \dvarepp{}).
We focused on two perspectives:
\begin{inparaenum}
    \item how effective, efficient, and robust the selected \glspl{llm} are in generating tests from medical rules;
    and
    \item how many potential issues we can identify in the implementation of \guri{} with our differential testing approach.
\end{inparaenum}

Our results showed that \glspl{llm} are highly effective in generating medical rule tests.
In particular, \gpt{} is the most effective \gls{llm} with hardly any hallucinations (i.e., mean completion rate of \perc{97.07050}) and being successful in generating \rulePass{} (\perc{82.10779}), \ruleFail{} (\perc{65.80779}), and \ruleNotApplied{} (\perc{79.19697}) tests.
However, \gpt{} is also the least efficient model with \s{15.496024} compared to the fastest models, i.e., \mixtral{} with \s{7.069893} and \mistral{} with \s{7.779271}.
In terms of robustness, \gpt{} performs the best for \rulePass{} tests with \perc{92.25741} and equally to \mixtral{} and \mistral{} for \ruleNotApplied{} tests with \perc{86.20417}.
Only for \ruleFail{} tests, \gpt{} with \perc{77.40914} is inferior to \mistral{} and \mixtral{} with both having \perc{86}.
Finally, regarding the differential testing results, we observe that \mixtral{} has the highest number of matches, while \mistral{} has the highest number of mismatches per rule. A total of 44 rule mismatches between \guri{} results and \dvarepp{} for \mistral{}, followed by \gpt{} with 32, \llama{} with 29 and \mixtral{} with 27.
While not all mismatches can be considered faults, our analysis reveals 22 rules that show inconsistencies in terms of handling either rule versions, date format, or variable dependencies.
These results suggest that \glspl{llm} are effective rule test generators to discover mismatches with differential testing, improving the current state of practice at the \gls{crn}.
Whether a mismatch is an actual fault still requires manual root cause analysis by the \gls{caress} developers.


\section{Real-World Application Context}
\label{sec:appcontext}



The \gls{crn} compiles information on cancer patients, covering data related to diagnosis and medical treatments
sourced from various healthcare organizations (e.g., hospitals and laboratories) as cancer messages.
The \gls{crn} created \gls{caress} to analyze and validate cancer messages and offer decision support to stakeholders like policymakers and researchers. 
The accuracy of the decisions made with \gls{caress} relies on the validity of the cancer messages it stores. 
To validate cancer messages, the \gls{crn} developed a subsystem of \gls{caress} named \guri{}---our application context. 
\guri{} is a web-based system that provides several \gls{rest} \gls{api} endpoints corresponding to various functionalities. 
One of the endpoints is the validation endpoint (\code{/api/messages/validation}) for validating cancer messages against medical rules. 

\begin{figure}[tbp]
\vspace{0.3em}
  \centering
    \includegraphics[width=\columnwidth]{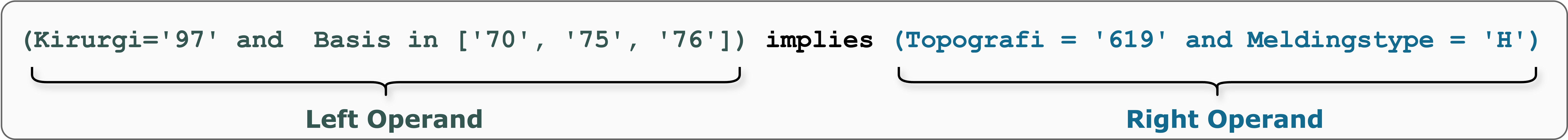}
    \caption{An example medical rule}
    \label{fig:rule}
\end{figure}

\dvare{}\footnote{\url{https://github.com/dvare/dvare-framework/}} is a rule-expression language engine that runs the rules on the data and asserts the results into \code{true} or \code{false} depending on if any of the condition matches.
\guri{} employs \dvare{} as the underlying engine to evaluate cancer messages against medical rules. 
While \dvare{} yields a Boolean result, \guri{} categorizes the results based on logical operations, particularly the \code{implies} operator.
Consider an example rule with two operands joined by an \code{implies} operation, as illustrated in \cref{fig:rule}. This rule can be described as follows:
If the surgical procedure (\code{Kirurgi}) is \code{'97'} and the basis code (\code{Basis}) is one of \code{'70'}, \code{'75'}, or \code{'76'}, then the topography code (\code{Topografi}) must be \code{'619'} and the message type (\code{Meldingstype}) must be \code{'H'}.
In this scenario, \guri{} handles the rules as follows:
\begin{inparaenum}
    \item If the \code{LeftOperand} is \code{true}, the rule is applied and can be either \rulePass{} or \ruleFail{} depending on the \code{RightOperand} (\code{true} for \rulePass{}, and \code{false} for \ruleFail{}).
    \item If the \code{LeftOperand} is \code{false}, the rule is \ruleNotApplied{}, irrespective of the \code{RightOperand}.
\end{inparaenum}
This categorization enables \guri{} to present rule validation outcomes with enhanced clarity for analysis and prediction within \gls{caress}.


Testing \guri{} is essential to ensure the validity of the cancer data and statistics produced by \gls{caress}~\cite{isaku:23}.
We identify four main challenges of testing \guri{} described in the following.
\textbf{(1) Evolution:} \guri{} and \dvare{} continuously evolve in response to changing cancer message formats and medical rules (i.e., rule deletion, modification, or insertion) dictated by standards/regulations or through cancer research. 
\textbf{(2) High-cost:} Testing multiple evolving implementations of \guri{} with various underlying environments incurs high time and resource costs. 
\textbf{(3) Labor-intensive:} Manual testing of each version \guri{} in different development phases is a laborious task.
Therefore, a cost-effective and automated approach is necessary to test the evolving \guri{}~\cite{laaber:23b}. \textbf{(4) Absence of Test Oracles:}
For medical rule testing, determining whether the rules are implemented correctly is largely manual, which presents significant challenges for automated testing. 
To address the absence of precise oracles, our approach uses differential testing to flag mismatches, supporting automated root cause analysis.

\section{Approach}
\label{sec:approach}

\begin{figure*}[tbp]
    \centering
    \includegraphics[width=0.8\textwidth]{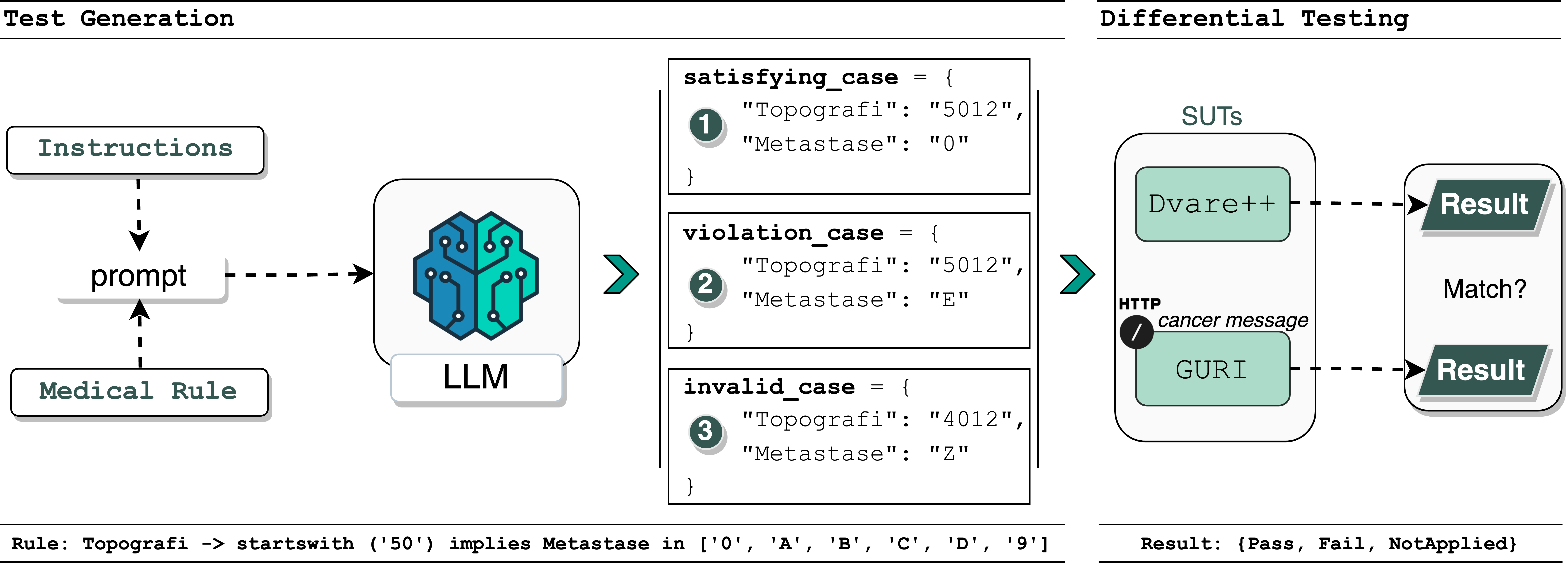}
    \caption{\ourApproach{} Overview}
    \label{fig:approach}
\end{figure*}

\begin{lstlisting}[float=tbp, style=prompt, caption=\Gls{llm} Prompt, label=lst:prompt, belowskip=0pt]
$$msg$$system_msg$$msg$$ $$python$$= f"""$$python$$
You are a cautious assistant, an expert in medical rules. To complete the user task, you need to strictly follow these steps:

$$step$$Step 1$$step$$: Clearly articulate the rule in natural language as you were explaining it to a child.
$$step$$Step 2$$step$$: Identify the variable names and generate values according to the rule's conditions.
- For satisfying the rule, provide a dictionary ('$$prop$$satisfying_case$$prop$$') including all variables specified in the rule.
- For violating the rule, alter ONLY the implied value while keeping other values unchanged; output a dictionary ('$$prop$$violation_case$$prop$$') for this.
- For cases that do not apply to the rule, provide a dictionary ('$$prop$$invalid_case$$prop$$') with invalid values that do not meet any rule condition.
$$step$$Step 3$$step$$: Double check the use of original variable names as properties, and maintain only one value for each variable. Refactor the dictionaries if needed.
$$step$$Step 4$$step$$: Express your confidence level on the generated test case using a score from 0% to 100%. Output '$$prop$$confidence_score$$prop$$' as a string.
$$step$$Step 5$$step$$: Format and output the data as a JSON object. Enclose all properties in double quotes.
$$python$$"""$$python$$

$$msg$$user_msg$$msg$$ $$python$$= f"""$$python$$
Identify the variables and values in the given rule and create a JSON object with the following properties: $$prop$$satisfying_case$$prop$$, $$prop$$violation_case$$prop$$, $$prop$$invalid_case$$prop$$, and $$prop$$confidence_score$$prop$$.

Rule: $$rule$${rule}$$rule$$
$$python$$"""$$python$$
\end{lstlisting}

\Cref{fig:approach} shows \ourApproach{}'s overview in two stages:
\begin{inparaenum}
    \item medical rule test generation
    and
    \item differential testing of \guri{}.
\end{inparaenum}

\paragraph{Test Generation}
This stage uses an \gls{llm}, which takes a prompt and a set of rules as input.
\ourApproach{} executes one prompt for each rule, resulting in three tests per rule: one \rulePass{}, one \ruleFail{}, and one \ruleNotApplied{} test.
A test has (only) the variable values that let the rule \rulePass{}, \ruleFail{}, or \ruleNotApplied{}.

\Cref{lst:prompt} shows the prompt divided into a system and user part.
In the system part (\colcode{msg}{system\_msg}), we set up the \gls{llm} to follow these steps:
\begin{description*}[afterlabel={{: }}, mode=unboxed, font=\colcode{step}]
    \item [Step 1] be precise;
    \item [Step 2] identify rule variables and generate tests:
    \begin{enumerate*}
        \item \rulePass{} (\colcode{prop}{satisfying\_case}),
        \item \ruleFail{} (\colcode{prop}{violating\_case}),
        and
        \item \ruleNotApplied{} (\colcode{prop}{invalid\_case});
    \end{enumerate*}
    \item [Step 3] do not be repetitive;
    \item [Step 4] express your confidence (\colcode{prop}{confidence\_score});
    and
    \item [Step 5] generate a \gls{json} object as the output.
\end{description*}

Note that we use different terms in the prompt, as the rules are similar to constraints in propositional logic and, hence, we expect the \gls{llm} to understand these terms better than the terms specific to \guri{}.
In the user part (\colcode{msg}{user\_msg}), we instruct the \gls{llm} to generate the three tests for a given rule, for which the prompt can be parameterized (\colcode{rule}{\{rule\}}).

\paragraph{Differential Testing}
In this stage, we execute the tests against two \glspl{sut}:
\begin{inparablank}
    \item \dvarepp{}
    and
    \item \guri{}.
\end{inparablank}
\dvarepp{} is a simplified reference implementation of \guri{} that internally uses the \dvare{} rule engine and adds a wrapper around it to return the same rule results as \guri{}, which requires minimal implementation and maintenance effort and ensures consistency with \guri{}, as \guri{} also uses \dvare{} as its underlying rule engine.
\guri{} is the real-world web-based rule engine of \gls{caress}, which validates and aggregates cancer messages through \gls{rest} endpoints.
As the generated tests only contain the variables of the rule they test, \ourApproach{} adds a preprocessing step before sending the tests to \guri{}, which embeds a test in a cancer message (formatted as \gls{json}) containing all the variables \guri{} requires.
For this, it uses a predefined set of values for the \enquote{other} variables not part of a test, which are randomly picked and do not yield an error.
These values remain the same across all the tests.
Differential testing solves the oracle problem~\citep{barr:15b} by sending the same input to two (or more) \glspl{sut} that should have the same outputs and comparing whether the actual outputs are identical~\citep{mckeeman:98}.
Automatically testing medical rule engines poses a challenge in generating test oracles, as the expected rule result for a given variable assignment is often unclear~\citep{laaber:23a}.
Therefore, leveraging differential testing provides a suitable solution to effectively test \guri{}.
For this, \ourApproach{} sends the same rule tests to \dvarepp{} and \guri{}, and checks whether the rule results match.
In case of a match, we consider the medical rule and the integration in \guri{} to be correct;
otherwise, we report to the \gls{crn} developers that there was a mismatch for a certain rule when running a certain test.

\section{Experimental Evaluation}
\label{sec:design}

To evaluate \ourApproach{}, we perform a laboratory experiment~\citep{stol:18} on \numllms{} \glspl{llm} and \numrules{} real-world medical rules from the \gls{crn}.
\subsection{Research Questions}
\label{sec:study:rqs}
Our experiment investigates the following \glspl{rq}.
\begin{description}[topsep=0.8em,labelindent=1em,labelwidth=3em,itemsep=0.2em]
    \item[RQ1] How effective and efficient are \glspl{llm} in generating tests from medical rules?
    \item[RQ2] How robust are \glspl{llm} when generating tests?
    \item[RQ3] To what extent can the generated tests be used for differential testing of a real-world medical rule engine?
\end{description}

\textbf{RQ1} studies the differences in the abilities of the selected \glspl{llm} in generating tests from three aspects:
\begin{inparaenum}
    \item hallucinations, i.e., syntactically invalid outputs;
    \item successful generations, i.e., semantically correct outputs (e.g., a test that should pass actually passes);
    and
    \item time efficiency.
\end{inparaenum}
%
\textbf{RQ2} focuses on the robustness of the selected \glspl{llm} to study their trustworthiness when generating tests for slightly mutated rules.
%
\textbf{RQ3} is designed to investigate the effectiveness of the generated tests in evaluating the \guri{} system, thereby studying the applicability of \ourApproach{} to a real-world system.


\subsection{\gls{llm} Selection and Settings}
\label{sec:study:llms}

We selected four \glspl{llm}:
\begin{inparaenum}
    \item Mistral 7B denoted as \mistral{}~\citep{mistral:23},
    \item Llama 2 13B denoted as \llama{}~\citep{llama:23},
    \item Mixtral 8x7b denoted as \mixtral{}~\citep{mixtral:24},
    and
    \item GPT-3.5 denoted as \gpt{}~\citep{gpt}.
\end{inparaenum}
We selected these models while considering availability, diversity, performance across general-purpose tasks, and popularity among the AI community.
\gpt{} from OpenAI, is a generative pre-trained transformer, while \llama{}, from Meta, is a pre-trained and fine-tuned \gls{llm} with 13 billion parameters.
We choose \mistral{} from Mistral AI for its good performance in reasoning, mathematics, and code generation, outperforming \llama{} on multiple benchmarks~\citep{mistral:23}.
Furthermore, \mixtral{}, a sparse mixture of expert models with open weights, distinguishes itself by matching or surpassing \gpt{} on various tasks~\citep{mixtral:24}.
Its impressive performance and widespread popularity within the AI community, particularly on platforms like HuggingFace were the reasons to select it.

We used the same temperature setting and number of repetitions for all the \glspl{llm}.
While different temperatures may impact the output of \glspl{llm}, specific guidelines and recommendations remain general; varying from 0 to 2, where higher values indicate more random or creative generation~\citep{temp:22}.
Although more empirical studies are needed, it is common to set the temperature around 0.7:
\begin{inparaenum}
    \item Llama 2~\citep{llama:23} uses 0.1 and 0.8 as defaults depending on the task and benchmark;
    \item the OpenAI \gls{api} reference defaults to 0.7;
    \item open-source frameworks such as The Rasa Community use 0.7;
    \item GTP-4 uses 0.6 as their \enquote{best-guess}~\citep{gpt-4};
    \item academic studies often use 0.7, aiming to balance determinism and creativity~\citep{juho:23,yifan:23,chung:23}.
\end{inparaenum}
Consequently, we employed 0.7 for all four models. To account for randomness, we let the \glspl{llm} generate tests for each rule \num{30} times~\citep{arcuri:11}.
Regarding the prompt (see \cref{sec:approach}), only \llama{} and \gpt{} support separate system and user parts.
For \mistral{} and \mixtral{}, we concatenate the two parts into a single prompt.
The three models \mistral{}, \llama{}, and \mixtral{} are accessed through the AI platform fireworks.ai, while \gpt{} is accessible only on the OpenAI platform.


\subsection{Medical Rules and Rule Mutations}
\label{sec:study:rules}

We selected a subset of the \gls{crn}'s rules, i.e., \num{70} validation rules, of which we
excluded \num{22} that contain complex structures, such as user-defined functions instead of variables, which would require access to the functions (as they are not part of the rule) and add complexity to \dvarepp{}.
In total, we use \originalRules{} rules in our study.

These rules exhibit varying levels of complexity across three aspects: clauses, variable names, and operators. The number of clauses ranges from 0 (representing rules with a single condition) to a maximum of 6. Similarly, the complexity of variable names varies from 1 (for simplistic rules with a single condition) to 7 (for the most complex rules), utilizing a total of 21 unique variable names. Furthermore, the employed operators range from 1 to 19, reflecting the diverse structural complexity of our rule dataset. For example, the rule in \cref{fig:rule} has 4 variables (i.e., Kirurgi, Basis, Topografi, and Meldingstype), which are literals in a logical expression connected by logical operators (e.g., \enquote{and}, \enquote{or}, \enquote{not}). In this case, the \enquote{implies} operator connects the antecedent and the consequent clause. So, the correct count of clauses in this case is 2. As for the number of operators, we have 7 in total, including both logical and relational operators.

In addition, to enlarge the studied rule set in RQ1 and compute the robustness metrics in RQ2, we define a set of mutation operators on medical rules (see \cref{tab:mutations}), inspired by previous research~\citep{shan:09,jia:11}.
The mutation operators consider
\begin{inparaenum}
    \item changing logical operators (\mutCO{}, \mutNI{}, and \mutACO{}), set containment (\mutRI{}), and built-in functions (\mutRSE{});
    \item swapping indices (\mutSSI{}) and rule clauses (\mutSR{});
    and
    \item altering dates (\mutACO{}).
\end{inparaenum}

We apply each mutation operator one at a time on every rule at every possible location.
Therefore, one mutation operator may be applied multiple times to the same rule when there are multiple possible locations, e.g., if a rule contains three \code{and} operators, we would apply \mutCO{} three times.
In total, we retrieve \mutatedRules{} mutated rules from the \originalRules{} rules. Note that we use the mutated rules only to evaluate the tests generated by the \glspl{llm} in RQ1 and RQ2 and not in RQ3, as \guri{} does not have these rules implemented and, hence, would not return the expected test results.

\input{mutation_types}

\subsection{Evaluation Metrics}
\label{sec:study:metrics}
\subsubsection{RQ1}
We follow previous works~\citep{em:1, em:2, em:3} and use EM to check if the completion (\gls{llm} test generation) matches exactly the expected output. 
An exact match is considered when the \gls{llm} can generate syntactically correct tests for all three types, i.e., \rulePass{}, \ruleFail{}, and \ruleNotApplied{}. 
This metric allows us to assess the \gls{llm}'s ability to produce valid tests, evaluating how accurately the \gls{llm} follows the instructions in the prompt.
Considering instruction inconsistency as an indicator of potential hallucination~\citep{huang:23}, we gain insight into the model's performance in handling different medical rules and maintaining consistency with the given instructions.
First, we introduce the completion rate metric in \cref{eq:completion_rate}, computed as the ratio of Exact Matches (EM) to the total expected cases ($T_{expected}$), representing the overall repetitions.
We use ${\#EM}$ to denote the number of times an \gls{llm} successfully generated valid tests.

\begin{equation}
    CR = \frac{\#EM}{T_{expected}}
    \label{eq:completion_rate}
\end{equation}

\input{rq1_metric}

\subsubsection{RQ2}
We assess the robustness of an \gls{llm} by introducing minor changes to each rule and studying how much the outputs of the original and mutated rules differ.
A similar way of assessing the robustness of \glspl{llm} has been employed by \citet{hyun:23}.
To this end, we calculate the average absolute difference between the success indices of each rule in the original and mutated sets for each test type.
The result is then subtracted from 1 to represent the robustness index, where a higher value indicates greater robustness for the individual rules within the test type.

\input{rq2_metric}

\subsubsection{RQ3}

\input{rq3_metric}


\subsection{Statistical Analyses}
\label{sec:stats}
We use hypothesis testing and effect sizes to statistically evaluate our result observations.
A set of observations is a distribution of metric values, e.g., by an \gls{llm}.
Depending on the \gls{rq}, we consider different observations analyzed by the statistical tests.
We follow the best practice~\citep{arcuri:11}.

We use the non-parametric Kruskal-Wallis H test~\citep{kruskal:52} to compare multiple observation sets, which is an extension of the pair-wise Mann-Whitney U test.
The null hypothesis $H_0$ states no statistical difference among the distribution sets' ranks.
The alternative hypothesis $H_1$ states that there is a difference.
If we can reject $H_0$ and accept $H_1$, we perform Dunn's post-hoc test~\citep{dunn:64} to identify for which pairs there is a difference.
We set the significance level to $\alpha = 0.01$ and control the false discovery rate with the Benjamini-Yekutieli procedure~\citep{benjamini:01} for multiple comparisons.

We additionally perform the Vargha-Delaney \vda{}~\citep{vargha:00} to assess the effect size magnitude.
Two observation sets are stochastically equivalent if \vda{}~$= 0.5$.
If \vda{}~$> 0.5$, the first observation is stochastically better than the second observation; otherwise \vda{}~$< 0.5$.
We further divide \vda{} into magnitude categories, based on $\hat{A}^{scaled}_{12} = (\hat{A}_{12} - 0.5) * 2$~\citep{hess:04}:
\begin{inparadesc}
    \item[negligible] $= |\hat{A}^{scaled}_{12}| < 0.147$,
    \item[small] $= 0.147 \leq |\hat{A}^{scaled}_{12}| < 0.33$,
    \item[medium] $= 0.33 \leq |\hat{A}^{scaled}_{12}| < 0.474$,
    and
    \item[large] $= |\hat{A}^{scaled}_{12}| \geq 0.474$.
\end{inparadesc} All results are considered statistically significant if the \pvalue{} is below $\alpha$ and the effect size is non-negligible.

\subsection{Threats to Validity}
\label{sec:threats}

\paragraph{Construct Validity}
The threats to construct validity are mostly concerned with the defined metrics.
We define effectiveness in terms of completion rate \metricCR{} and success index \metricSI{}, and efficiency with the inference time \metricIT{}.
Our robustness metric \metricRI{} builds on rule mutations.
Hence, the choice of mutation operators, which is based on previous work~\citep{shan:09,jia:11}, is key to the validity of \metricRI{}.
Other metrics might lead to different results.
\metricIT{} includes the \gls{http} request sent to the \gls{llm} platform (i.e., OpenAI or Fireworks.ai) and not only the model inference time.
However, the \gls{http} time should be similar across the models, as we use the same platform for \llama{}, \mistral{}, and \mixtral{}, only for \gpt{} we use OpenAI's \gls{api}.

\paragraph{Internal Validity}
Internal validity concerns the experiment setup. 
We used hosted \glspl{llm}, limiting our control over model execution. 
While we cannot guarantee that model instances or sessions are not shared across users or \gls{http} requests, platform documentation suggests each request is treated independently, minimizing leakage risks. 
We maintained consistent \gls{llm} temperature settings, though results may vary with different values. 
To mitigate stochastic effects, we followed best practices, repeating experiments \num{30} times and using statistical tests~\citep{arcuri:11}. 
Despite the inherent possibility of bugs in the pipeline, we reviewed scripts and reran experiments to avoid critical errors.

\paragraph{External Validity}
Threats to external validity relate to the generalizability of our results, in particular, beyond the \glspl{llm}, medical rules, mutation operators, and the rule engine \guri{}.
We select a representative set of \glspl{llm} that have shown strong performance in general-purpose tasks and widespread popularity in the AI community.
In terms of rules, we select a subset of \guri{}'s validation rules.
Our results might change for different validation rules, aggregation rules, or different rules from other cancer registries, potentially from other countries.
Finally, we perform a laboratory experiment, i.e., we run the tests against a standalone version of \guri{} in an isolated setting and do not perform a field experiment, where the tests are executed against the real \guri{}.

\section{Results and Analyses}
\label{sec:results}
\subsection{RQ1: Effectiveness and Efficiency}
\label{sec:results:rq1}

This section investigates the completion rate \metricCR{}, success index \metricSI{}, and inference time \metricIT{} across the \glspl{llm}.

\begin{figure}[tbp]
    \centering
        \includegraphics[width=\columnwidth]{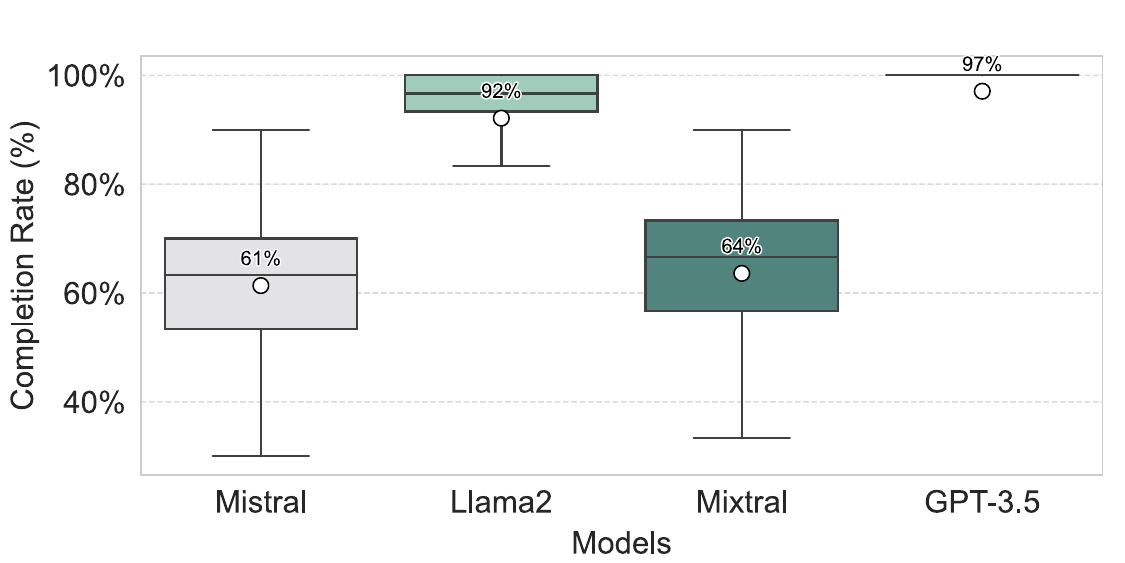}
    \caption{Completion rates for all the rules per \gls{llm}. 
    }
    \label{fig:completion_rate}
\end{figure}

\input{rq1_table_stats.tex}

\paragraph{Completion Rate}
\Cref{fig:completion_rate} depicts the \metricCR{} across the \glspl{llm}.
We observe that \gpt{} performs the best with a mean \metricCR{} of \perc{97.07050}, indicating that it almost never hallucinates and always generates the correct \gls{json} test output.
The second-best model is \llama{} with \perc{92.12339}, before there is a considerable drop for \mixtral{} and \mistral{} to \perc{63.60529} and \perc{61.35958}, respectively.
The statistical tests, depicted in \cref{tab:rq1:stats}, support these observations: \gpt{} is statistically better than \llama{} with medium and \mixtral{} and \mistral{} with large effect sizes.

\begin{figure}[tbp]
    \centering
        \includegraphics[width=\columnwidth]{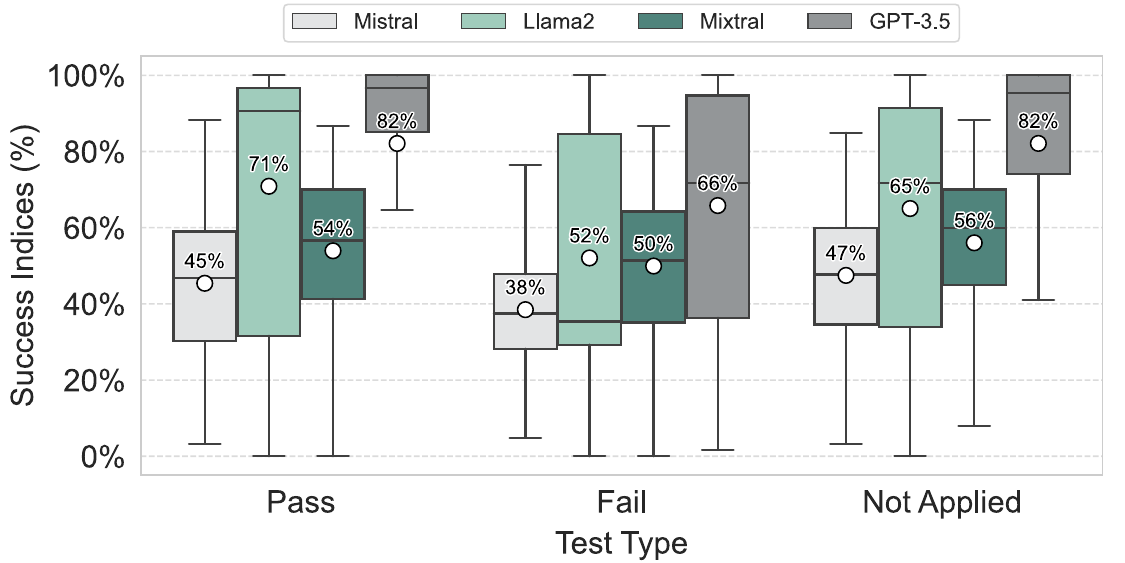}
    \caption{Success indices for all the rules per test type and \gls{llm}.}
    \label{fig:success_rates}
\end{figure}

\paragraph{Success Index}
\Cref{fig:success_rates} shows the success indices \metricSIPass{}, \metricSIFail{}, \metricSINA{} for the \rulePass{}, \ruleFail{}, and \ruleNotApplied{} tests, respectively, across the \glspl{llm}.
Similar to the completion rates, we observe that \gpt{} performs the best, with a mean \metricSIPass{} of \perc{82.10779}, \metricSIFail{} of \perc{65.80779}, and \metricSINA{} of \perc{82.12125},
followed by \llama{}, \mixtral{}, and \mistral{}.
The statistical tests confirm \gpt{}'s superiority: for all three test types, it is better than all the other \glspl{llm}, at least with a small effect size.
While the \glspl{llm} perform similarly on \metricSIPass{} and \metricSINA{}, they exhibit a considerable drop on \metricSIFail{}, except \mixtral{} which performs similarly across the three metrics.
This shows that the \glspl{llm} struggle more to generate \ruleFail{} tests, which might be due to the structure of the rules, where the \gls{llm} must satisfy the left clause of the implication while it must falsify the right clause (see \cref{sec:appcontext}).

\begin{figure}[tbp]
    \centering
        \includegraphics[width=\columnwidth]{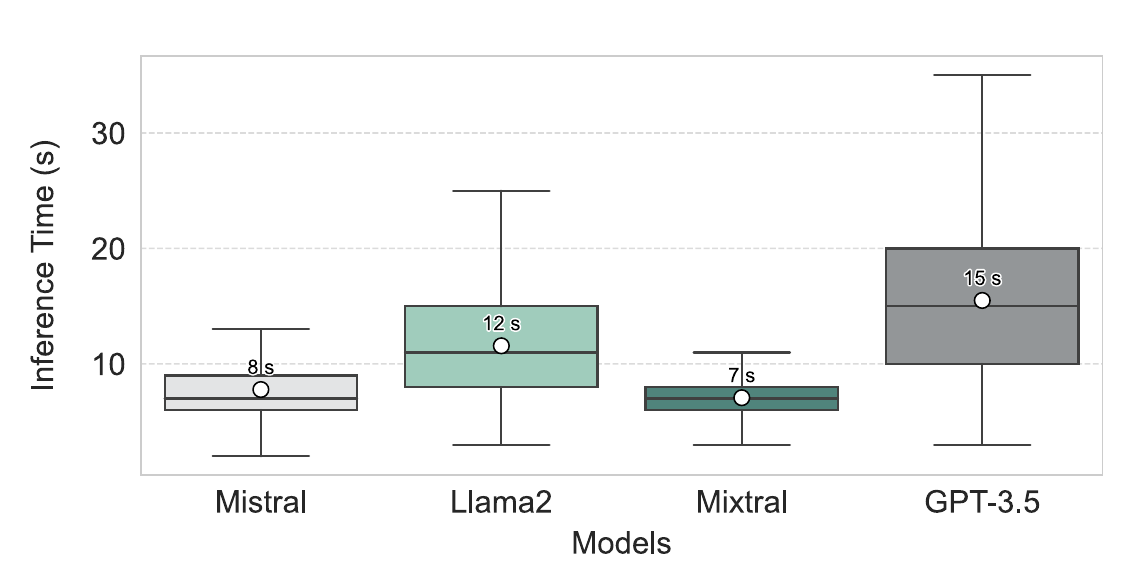}
    \caption{Inference times for all the rules per \gls{llm}.}
    \label{fig:inference_time}
\end{figure}

\paragraph{Inference Time}
In terms of efficiency, \cref{fig:inference_time} plots the inference times \metricIT{} for all the rules across the \glspl{llm}.
We observe that the most effective \gls{llm}, i.e., \gpt{}, is the least efficient one, with a mean \metricIT{} of \s{15.496024}.
The fastest \glspl{llm} are \mixtral{} with \s{7.069893} and \mistral{} with \s{7.779271}, followed by \llama{} with \s{11.568463}.
In terms of statistical tests, \mixtral{} and \mistral{} are equal, while \gpt{} is outperformed by \mixtral{} and \mistral{} with large and by \llama{} with small effect size.
This shows that higher effectiveness comes at a considerable runtime cost.


\begin{Summary}
    \textbf{RQ1 Summary:} 
    \gpt{} is the most effective \gls{llm} for generating medical rule tests; however, it is also the least efficient one, where \mixtral{} and \mistral{} are the most efficient ones.
\end{Summary}

\subsection{RQ2: Robustness}
\label{sec:results:rq2}

\begin{figure}[tbp]
    \centering
        \includegraphics[width=\columnwidth]{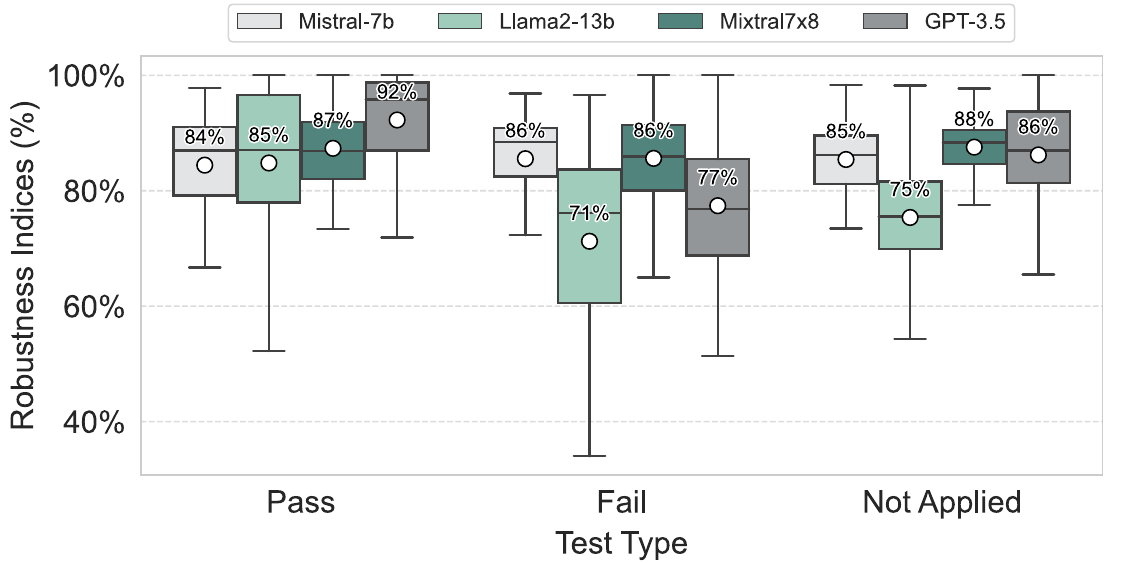}
    \caption{Robustness indices for all the rules per test type and \gls{llm}.}
    \label{fig:robustness}
\end{figure}

\input{rq2_table_stats_rules.tex}

This section investigates the \gls{llm} robustness \metricRI{} per rule and test type.
%
%
\Cref{fig:robustness} depicts the robustness indices \metricRIPass{}, \metricRIFail{}, and \metricRINA{} for the \rulePass{}, \ruleFail{}, and \ruleNotApplied{} tests per rule, respectively.
The results are more diverse compared to the success indices: no \gls{llm} is superior across the three test types.
In terms of \metricRIPass{}, \gpt{} is most robust with a mean of \perc{92.25741}, followed by \mixtral{}, \llama{}, and \mistral{} with \perc{87.34621}, \perc{84.80741}, and \perc{84.43345}, respectively.
This is confirmed by the statistical tests (see \cref{tab:rq2:stats:rules}): \gpt{} is better than \mixtral{} and \llama{} with medium and \mistral{} with large effect size.
For \metricRIFail{}, \mistral{} and \mixtral{} tied for the most robust \gls{llm} with \perc{86}, followed by \gpt{} with \perc{77.40914} and a medium effect size and \llama{} with \perc{71.24534} and a large effect size.
Finally, for \metricRINA{}, we do not observe statistical differences between \mistral{}, \mixtral{}, and \gpt{} with \perc{85.41750}, \perc{87.55125}, and \perc{86.20417}, respectively.
Only \llama{} is inferior to the other three \glspl{llm} with \perc{75.36792} and large effect sizes.

\begin{Summary}
    \textbf{RQ2 Summary:} 
    \gpt{} is the most robust \gls{llm} when generating \rulePass{} and \ruleNotApplied{} tests, but is less robust than \mistral{} and \mixtral{} for \ruleFail{} tests.
\end{Summary}

\subsection{RQ3: Differential Testing}
\label{sec:results:rq3}
We investigate matches and mismatches between \dvarepp{} and \guri{}.
In terms of mismatches, we categorize them based on the result type (i.e., \rulePass{}, \ruleFail{}, \ruleNotApplied{}, and \ruleWarning{}) and \enquote{errors} returned by \guri{}.
A \ruleWarning{} result occurs when the rule execution is successful, but the data exhibits potential issues. For example, a patient's age might be 120 years, which is theoretically possible but highly unlikely; hence the \ruleWarning{} result. In addition, we identify two other result types:
\begin{inparaenum}
    \item \guriRule{500}---thrown as an exception by \guri{} in response to invalid or non-compliant variables (e.g., date data type), deviating from the specific implementation standards. For instance, if a date is provided in a format other than expected (e.g., \code{YYYY-MM-DD}), it triggers a 500 Internal Server Error.
    \item \guriRule{Empty Response}---similar to the above error, an empty response is triggered in specific instances where certain variables dictate conditions or requirements for the rule's execution.
    For example, if the diagnosis' certainty, identified with the \emph{\enquote{ds}} variable, is either invalid or null in a cancer message, the system generates an empty response due to pre-aggregation constraints. 
\end{inparaenum}
We have six categories based on the \guri{} result types.
A mapping between \guri{} with \dvarepp{} results is done to determine if a match or mismatch is observed.
\Cref{tab:rq3_results} shows all the observed matches and mismatches after executing each generated test.

We notice that \mixtral{} has the highest number of matches, while \mistral{} has the highest number of mismatches per rule. A total of 44 rule mismatches between the \guri{} and \dvarepp{} results for \mistral{}, followed by \gpt{} with 32, \llama{} with 29, and \mixtral{} with 27.
In addition, we perform a manual investigation of the nature of the mismatched rules and find the following patterns:
\begin{inparaenum}
    \item \textit{Rules that contain pre-aggregation constraints.} We observe that all 11 rules containing the \enquote{ds} variable result in an \guriRule{Empty Response} when its value is invalid or missing.
    \item \textit{Rules that return \guriRule{500}.} We encounter two rules, specifically V47 and V53, that trigger a \guriRule{500} result due to either an invalid date or a non-compliant date format. 
    \item \textit{Rules that return \ruleWarning{}.} Among all the 58 rules, we observe only one rule, i.e., V08, which returns \ruleWarning{} instead of \ruleFail{}.
    \item \textit{Rules that always return \ruleNotApplied{}.} We observe that the rules V43, V44, and V45 always return \ruleNotApplied{}, regardless of the test type (i.e., \rulePass{}, \ruleFail{}, \ruleNotApplied{}).
    \item \textit{Rules that never \rulePass{}.} We observe a list of rules where older versions can never \rulePass{}. Specifically, V19/1, V20/1, V22/1, V27/1 (resulting always in \ruleFail{}) and V69/1 (resulting always in \ruleNotApplied{}). 
\end{inparaenum}

Based on the identified result mismatches, we observe the following reasons why such mismatches can occur:
\begin{inparaenum}
    \item \textit{Inconsistency in handling rule versions.} We notice that some rules have both versions \enquote{active} (e.g., V09/1 and V09/2), while some other rules are \enquote{inactive} for the first version. Thus, such rules never \rulePass{}.
    \item \textit{Inconsistency in handling date formats.} We notice that \guri{} considers \code{YYYY-MM-DD} as a valid date format, while \code{DD-MM-YYYY} is invalid, thus leading to a \guriRule{500} result.
%
    \item \textit{Inconsistent variable handling.} Different variables are handled differently in terms of execution. For instance, \enquote{ds} affects the whole cancer message (resulting in empty response), while other variables (e.g., diagnose date) only impact specific rules (always resulting in \ruleNotApplied{}).
\end{inparaenum}

\input{rq3_table}



\begin{Summary}
    \textbf{RQ3 Summary:} 
    \mistral{} exhibits the most mismatches per rule, followed by \gpt{}, \llama{}, and \mixtral{}.
    While not all mismatches can be considered faults, we observe 22 rules that show inconsistencies in terms of handling either rule versions, date format, or variable dependencies. 
\end{Summary}

\section{Discussion and Lessons Learned}
\label{sec:lessons}

\subsection{Hallucination Effect} 
As an indicator of potential hallucination, we consider the \glspl{llm} accuracy in following the given prompt instructions. 
We utilize the $EM$ metric to determine whether the tests generated by the \glspl{llm} are correct. 
Despite its common use, we identify several limitations:
\begin{inparaenum}
    \item \textit{Semantic alterations.} This limitation refers to small semantic changes, such as transforming \enquote{violation\_case} to \enquote{violating\_case}, that lead to incorrect outputs.
    \item \textit{Different levels of detail.} We observe three discrepancies between the expected and generated tests:
    \begin{inparaenum}
        \item Missing test types: In some instances, the \glspl{llm} only generate a subset of the three specified test types (i.e., \rulePass{}, \ruleFail{}, and \ruleNotApplied{}).
        \item Additional tests: The \glspl{llm} produce redundant tests, unnecessarily generating variations of the same scenario (e.g., generating \enquote{invalid\_case}, \enquote{invalid\_case\_1}, and \enquote{invalid\_case\_2}).
        \item Lack of integration: Tests are generated individually and not as a single \gls{json} dictionary containing all three types. 
    \end{inparaenum}
    \item \textit{Invalid \gls{json} format.} The \gls{json} parser throws exceptions for issues including:
    \begin{inparaenum}
        \item Improperly formatted property names: \gls{json} requires property names to be enclosed in double quotes.
        \item Missing variable and value pairs: Errors can occur when the \gls{json} misses information, such as variable names or their corresponding values.
        \item Incorrect data structure: Instead of generating the expected dictionary format (key-value pairs), the \gls{llm} generates lists.
        \item Missing delimiters: \gls{json} parsing errors also occur when expected delimiters, e.g., \enquote{:} or \enquote{,}, are missing. 
    \end{inparaenum}
\end{inparaenum}



\subsection{No \gls{llm} Subsumes Another}
Our differential testing results reveal that the most effective \gls{llm} (i.e., \gpt{}) does not uncover the most mismatches. 
Surprisingly, it is \mistral{} that detects the highest number of mismatches.
However, it is important to note that not all mismatches indicate faults. 
We identify three specific cases leading to inconsistencies when handling
\begin{inparaenum}
    \item \textit{rule versions,}
    \item \textit{date formats,}
    and
    \item \textit{variable dependencies.}
\end{inparaenum}
While 22 rules fall into these categories, none of the \glspl{llm} manages to identify all of them.
For instance, while \mistral{} successfully detects all rules involving the \enquote{ds} variable, it misses cases triggering a 500 status code---unlike \mixtral{}.
Contrarily, \mixtral{} identifies both cases triggering a 500 status code but misses those where all the rule versions (e.g., V20/1) never \rulePass{}. 
Interestingly, both \llama{} and \gpt{} detect all the cases related to rule versions but fail to uncover the cases where \mistral{} and \mixtral{} excel.
While any of the \glspl{llm} would be suited for detecting mismatches, the optimal choice requires a deeper understanding of the specific inconsistency types each model excels at identifying and the concrete usage scenario.
Additionally, exploring other \glspl{llm} might improve detecting specific inconsistencies or offer novel insights.

\subsection{Using \glspl{llm} Instead of Other Test Generation Techniques}
A specialized testing approach is essential for ensuring the correctness of the medical rules and their implementation in \guri{}.
While common testing methods like random and coverage-guided testing are widely used~\citep{hamlet:21, hong:97}, they fall short of capturing the specifics of medical rules, such as rule context and relationships among variables.
\ourApproach{} introduces a customized, domain-aware strategy that leverages \glspl{llm} to generate tests tackling the unique aspects of medical rules and \guri{} (e.g., \gls{icd} codes).
Moreover, as the medical rules are similar to constraints, a test generation technique might rely on constraint solvers to find suitable tests.
While such solvers are efficient in finding satisfying and violating variable assignments, they may not capture the \enquote{meaning} of variable names and relationships among variables and constraints. 
This capability is not limited to assigning appropriate values to variables. 
It also ensures that the generated test inputs consider the logical and domain-specific constraints between variables, such as temporal dependencies (e.g., diagnosis date preceding treatment date) or valid combinations of attributes (e.g., certain diagnoses requiring specific procedures).
Nonetheless, a dedicated empirical evaluation is needed to compare \glspl{llm} with other constraint-solving approaches (e.g., based on search algorithms and \gls{smt} solvers).   
\subsection{\Glspl{llm} Might Not Find Simple Faults}
The power of \glspl{llm} is that they can interpret the variable names to generate valid values.
For dates this can be helpful, as a date string needs to follow a specific format (e.g., \code{YYYY-MM-DD}); otherwise, the rule engine will yield a parsing error.
An \gls{llm} will always generate a date string of a valid format; however, it might never try a simple input such as a single number (e.g., \code{7}).
This specific case would be detected in our differential testing setting, where \dvarepp{} yields an error but \guri{} accepts a single number as a date variable, but the \gls{llm} never generates such a simple case.
A simpler test generator, e.g., a fuzzer like \tool{EvoMaster}~\citep{arcuri:19}, would likely try such an input and detect the fault.

\subsection{Mismatches $\neq$ Faults}
In RQ3, we analyzed differential testing results and demonstrated that \ourApproach{} effectively uncovers disparities within rule engines. 
However, its role is limited to flagging mismatches, which may include false positives and negatives (e.g., data overlaps between rule versions leading to \rulePass{}). 
Mismatches are not inherently good or bad; their value lies in their ability to provide actionable insights. 
Useful mismatches highlight faults in rule logic or system behavior, while non-actionable mismatches, such as those caused by invalid input formats, indicate areas for improving input validation or test generation. 
While more mismatches may suggest better test coverage, they also increase the workload for manual analysis, requiring developers to discern actionable issues. 
Thus, the focus should not be on maximizing or minimizing mismatches but on generating meaningful tests that reveal actionable insights and improve system reliability. 
Over time, the reference implementation can be adapted to handle edge cases that previously led to false positives.
Alternatively, it is worth investigating whether learned models can serve as a reference implementation, e.g., a digital twin of \guri{}~\citep{chengjie:23}.

\subsection{Improvement on the State of the Practice}
The rule testing process at the \gls{crn} is currently done manually by medical coders, who run a set of tests against new or changed rules whenever \enquote{necessary}~\citep{laaber:23b}.
Instead, automated rule test generation would relieve them of this tedious task.
Unfortunately, and to the best of our knowledge, there are no automatic test generation techniques for medical rule engines that have a solution for the oracle problem~\citep{barr:15b}.
In our previous works, we explicitly pointed out that generating valid medical data~\citep{laaber:23a}, solving the oracle problem for medical rules and rule engines~\citep{laaber:23a}, and keeping the number of tests executed against the real system to a minimum~\citep{isaku:23} is detrimental for being applicable at the \gls{crn}.
In this paper, we address these challenges by generating only three valid and realistic tests, i.e., \rulePass{}, \ruleFail{}, and \ruleNotApplied{}, for each rule and using differential testing with a simple reference implementation, i.e., \dvarepp{}.

\subsection{Generality of the Results}
While the results are specific to the \gls{crn}, \gls{caress}, and \guri{}; the medical rules used in the experiment; and the studied \glspl{llm}; we believe that \ourApproach{} and the overall findings are transferable to other contexts.
\begin{inparaenum}
    \item Our findings are relevant to other health registries in Norway (e.g., the Norwegian Patient Registry) and similar real-world contexts, such as the Norwegian toll and customs, which implement similar rule-based systems.   
    \item Other countries also maintain health registries, such as cancer registries, for which our findings and lessons can provide valuable guidance for adopting and applying an \gls{llm}-based test generation approach such as \ourApproach{}.
    \item Our medical rules are an instance of logic-based rules or constraints; therefore, any system taking such rules as input can leverage an approach like \ourApproach{}.
    \item The differential testing approach of \ourApproach{} uses a simple reference implementation to compare against, as alternative implementations of highly customized systems, such as \guri{}, are unlikely to exist.
    Hence, systems with similar settings aiming to test beyond simple crash oracles can follow a testing approach like ours.
\end{inparaenum}

\section{Related Work}
\label{sec:relatedwork}

A prominent software testing challenge is the absence of precise and reliable test oracles~\citep{tsong:18}, i.e., checking a test case's actual outcome with the expected outcome using a test oracle. In many real-world situations (including our context), such an oracle either doesn't exist or is impractical due to resource constraints.
Metamorphic and differential testing techniques are solutions to the test oracle problem~\citep{segura:18, godefroid:20}.
In this paper, we applied differential testing to \guri{} using \dvarepp{} as a simple reference implementation.
The idea is to check whether both systems produce consistent outputs on identical inputs generated by \glspl{llm} and identify inconsistencies stemming from potential implementation issues of \guri{}.
As a result, we address the test oracle problem to some extent. 

Recently, \glspl{llm} have gained popularity in software testing~\citep{wang:23}, including unit testing~\citep{vitor:23, xie:23}, system input generation~\citep{deng:23, moradi:23}, property-based test generation~\citep{vasudev:23}, and test oracle generation~\citep{tufano:22}. Our work focuses on generating tests with \glspl{llm} to produce \rulePass{}, \ruleFail{}, or \ruleNotApplied{} results, specifically for the medical domain. Related work includes \citet{zhang:23}, who evaluate \glspl{llm}, such as GPT-4 and Llama 2, for medical term classification using metastatic cancer data, and \citet{xu:23}, who assess models like GPT-4 on clinical text, emphasizing the need for better evaluation aligned with healthcare standards. Surveys~\citep{peng:23, thirunavukarasu:23, clusmann:23} highlight the potential and challenges of \glspl{llm} in healthcare, noting limited practical use. Our work contributes to advancing the practical deployment of \glspl{llm} for testing \gls{crn} software systems.


In the past, we developed a model-based framework to model the medical rules for \guri{}~\citep{wang:16} with the \gls{uml} and \gls{ocl}.
Next, we developed an impact analysis approach~\cite {wang:17} to study the impact of rule changes in response to \guri{}'s evolution (e.g., changes in regulations).
In addition, we developed a search-based approach to automatically refactor rules~\cite{lu:19} to enhance rule understandability and maintainability.
However, these approaches do not help find problems in the modeled rules.
More recently, four studies advanced the testing at the \gls{crn}, mainly focusing on the \guri{} subsystem~\citep{laaber:23a, laaber:23b, isaku:23, chengjie:23}. 
\citet{laaber:23b} address the practical complexities of testing an evolving system, offering insights applicable to healthcare registries.
Moreover, we studied EvoMaster~\citep{arcuri:19} (an AI-based testing tool) to assess its effectiveness regarding code coverage, error detection, and domain-specific rule coverage at the \gls{crn}~\citep{laaber:23a}. 
We extended EvoMaster and proposed EvoClass~\citep{isaku:23} which uses a \gls{ml} classifier to significantly reduce cost without compromising effectiveness.
Lastly, EvoCLINICAL~\citep{chengjie:23}, a cyber-cyber digital twin for \guri{}, was proposed, which with fine-tuning through active transfer learning, can deal with system evolution.
In contrast to these works, this paper addresses the test oracle problem associated with rule execution results by employing differential testing.
Moreover, we tackle the challenge of generating \glspl{ehr} by leveraging the capabilities of \glspl{llm}.

\section{Conclusions}
\label{sec:conclusions}
This paper presented \ourApproach{}, an \gls{llm}-based test generation approach for medical rules employed to perform differential testing of \guri{}, the \gls{crn}'s medical rule web service.
While we find that \ourApproach{} is effective in generating tests and finding mismatches, there are a few future directions worth investigating:
\begin{inparaenum}
    \item generate functions with \glspl{llm} that produce \rulePass{}, \ruleFail{}, and \ruleNotApplied{} tests;
    \item using \glspl{llm} to generate rule mutations;
    \item fine-tune an \gls{llm} with domain-specific \glspl{ehr} and \gls{crn} data;
    and
    \item combine other test generation techniques with \ourApproach{}.
\end{inparaenum}

\section*{Data Availability}
Due to confidentiality, we can not disclose the data, scripts, and systems that lead to the results of the paper.

\section*{Acknowledgments}

The Norwegian Ministry of Education and Research supports Erblin Isaku's PhD work reported in this paper. The research presented is also partly supported by \gls{rcn} under the project \texttt{309642} and has benefited from the \glsentryfull{ex3}, which is supported by the \gls{rcn} project \texttt{270053}.

\footnotesize
\bibliographystyle{IEEEtranSN}
\bibliography{refs}

\end{document}

%% file: packages.tex
\usepackage[T1]{fontenc}
\usepackage[utf8]{inputenc}
\usepackage[american]{babel}
\usepackage{csquotes}
\usepackage{microtype}
\usepackage[defblank]{paralist}
\setdefaultenum{(1)}{(a)}{(i)}{A.}
\usepackage[inline]{enumitem}
\usepackage{url}
\usepackage{flushend}
\usepackage{etoolbox}
\usepackage{fix-cm}
\usepackage{textpos}
\usepackage{mfirstuc}
\usepackage{titlecaps}
\usepackage[datesep=.,style=ddmmyyyy]{datetime2}

\usepackage{algorithm}
\usepackage{algpseudocode}

\usepackage{amsmath}
\usepackage{amssymb}
\usepackage{amsfonts}
\usepackage{bm}
\usepackage{relsize}
\usepackage{siunitx}
\usepackage[super]{nth}

\usepackage{booktabs}
\usepackage{multirow}
\usepackage{colortbl}
\usepackage{makecell}
\usepackage{rotating}
\usepackage{threeparttable}

\usepackage{float}
\usepackage{graphicx}
\usepackage[table,dvipsnames]{xcolor}
\usepackage[most]{tcolorbox}

\usepackage{listingsutf8}

\usepackage{xparse}
\usepackage{xspace}

\usepackage[xindy,acronym]{glossaries}
\glsdisablehyper

\usepackage[numbers,sort&compress]{natbib}

\usepackage[hidelinks,bookmarks=false]{hyperref}
\usepackage[all]{hypcap}

\usepackage[capitalise]{cleveref}

%% file: acronyms.tex
\newacronym{api}{API}{application programming interface}
\newacronym{caress}{CaReSS}{cancer registration support system}
\newacronym{crn}{CRN}{Cancer Registry of Norway}
\newacronym{ehr}{EHR}{electronic health record}
\newacronym{ex3}{eX\textsuperscript{3}}{Experimental Infrastructure for Exploration of Exascale Computing}
\newacronym{http}{HTTP}{Hypertext Transfer Protocol}
\newacronym{icd}{ICD}{International Classification of Diseases}
\newacronym{json}{JSON}{JavaScript Object Notation}
\newacronym{llm}{LLM}{large language model}
\newacronym{ml}{ML}{machine learning}
\newacronym{niph}{NIPH}{National Institute of Public Health}
\newacronym{ocl}{OCL}{Object Constraint Language}
\newacronym{rcn}{RCN}{The Research Council of Norway}
\newacronym{rest}{REST}{representational state transfer}
\newacronym{rq}{RQ}{research question}
\newacronym{smt}{SMT}{satisfiability modulo theories}
\newacronym{sut}{SUT}{system under test}
\newacronym{uml}{UML}{Unified Modeling Language}

%% file: commands.tex
\newcommand{\s}[1]{\SI{#1}{\second}}
\newcommand{\perc}[1]{\num[round-precision=0]{#1}\%}
\newcommand{\pval}[1]{\num[round-precision=5, round-pad=false]{#1}}
\newcommand{\vdaval}[1]{\num[round-precision=3, round-pad=true]{#1}}

\newcommand{\pvalue}{\textit{p}-value}
\newcommand{\vda}{$\hat{A}_{12}$}
\newcommand{\colcode}[2]{{\color{#1}\texttt{#2}}} 
\newcommand{\code}[1]{\colcode{black}{#1}}
\newcommand{\tool}[1]{\textit{#1}}

\newcommand{\numllms}{four}

\newcommand{\ourApproach}{LLMeDiff}
\newcommand{\dvare}{\tool{Dvare}}
\newcommand{\dvarepp}{\tool{Dvare++}}
\newcommand{\guri}{\tool{GURI}}

\newcommand{\guriRule}[1]{\textit{#1}}
\newcommand{\rulePass}{\guriRule{Pass}}
\newcommand{\ruleFail}{\guriRule{Fail}}
\newcommand{\ruleNotApplied}{\guriRule{NotApplied}}
\newcommand{\ruleWarning}{\guriRule{Warning}}
\newcommand{\originalRules}{\num{58}}
\newcommand{\numrules}{\originalRules}
\newcommand{\mutatedRules}{\num{322}}

\newcommand{\llm}[1]{\textit{#1}}
\newcommand{\mistral}{\llm{Mistral}}
\newcommand{\mixtral}{\llm{Mixtral}}
\newcommand{\llama}{\llm{Llama2}}
\newcommand{\gpt}{\llm{GPT-3.5}}

\newcommand{\metricCR}{$CR$}
\newcommand{\metricIT}{$t_{infer}$}
\NewDocumentCommand{\metricSI}{o}{%
    \IfValueTF{#1}
        {$SI_{#1}$}
        {$SI$}%
}
\newcommand{\metricSIPass}{\metricSI[\rulePass]}
\newcommand{\metricSIFail}{\metricSI[\ruleFail]}
\newcommand{\metricSINA}{\metricSI[\ruleNotApplied]}
\NewDocumentCommand{\metricRI}{o}{%
    \IfValueTF{#1}
        {$RI_{#1}$}
        {$RI$}%
}
\newcommand{\metricRIPass}{\metricRI[\rulePass]}
\newcommand{\metricRIFail}{\metricRI[\ruleFail]}
\newcommand{\metricRINA}{\metricRI[\ruleNotApplied]}

\newcommand{\mut}[1]{\textit{#1}}
\newcommand{\mutACO}{\mut{ACO}}
\newcommand{\mutAD}{\mut{AD}}
\newcommand{\mutCO}{\mut{CO}}
\newcommand{\mutNI}{\mut{NI}}
\newcommand{\mutRSE}{\mut{RSE}}
\newcommand{\mutRI}{\mut{RI}}
\newcommand{\mutSR}{\mut{SR}}
\newcommand{\mutSSI}{\mut{SSI}}

\newcommand{\statWorse}{worse}
\newcommand{\statBetter}{better}
\newcommand{\statEqual}{equal}

\definecolor{greylight}{RGB}{240,240,240}
\definecolor{greymedium}{RGB}{189,189,189}
\definecolor{greydark}{RGB}{99,99,99}

\definecolor{msg}{RGB}{0,0,128}
\definecolor{python}{RGB}{150,150,150}
\definecolor{step}{RGB}{128,128,0}
\definecolor{prop}{RGB}{102,14,122}
\definecolor{rule}{RGB}{32,153,157}

\lstdefinestyle{prompt}{ %
    moredelim=[is][\color{msg}]{$$msg$$}{$$msg$$},
    moredelim=[is][\color{python}]{$$python$$}{$$python$$},
	moredelim=[is][\color{step}]{$$step$$}{$$step$$},
	moredelim=[is][\color{prop}]{$$prop$$}{$$prop$$},
    moredelim=[is][\color{rule}]{$$rule$$}{$$rule$$},
	numberstyle=\tiny\color{black},
	showstringspaces=false,
	backgroundcolor=\color{white},				
	basicstyle=\fontsize{7}{8}\ttfamily,		
	breakatwhitespace=false,   					
	breaklines=true,							
    breakindent=1em,                            
	captionpos=b,								
	extendedchars=true,							
	frame=single,								
	keepspaces=true,							
	keywordstyle=\color{keywords}\bfseries,		
	numbers=left,								
	numbersep=10pt,								
	rulecolor=\color{black},					
	showspaces=false,							
	showstringspaces=false,						
	showtabs=false,								
	tabsize=4,									
	title=\lstname,								
	stepnumber=1,
	firstnumber=1,
	xleftmargin=2em,
	xrightmargin=2em,
	framexbottommargin=5pt,
	framextopmargin=5pt,
	framexleftmargin=2em,
	framexrightmargin=2em,
}

%% file: mutation_types.tex
\begin{table}[tbp]
    \centering
    \scriptsize
    \caption{Mutation operators}
    \label{tab:mutations}
    \begin{tabular}{llp{18em}}
        \toprule
        \textbf{Name} & ID & \textbf{Description} \\
        \midrule
        Alter Comparison Operators & \mutACO{} & Modify \code{>} to \code{<} and \code{<=} to \code{>=} and vice versa. \\
        Alter Date & \mutAD{} & Alter the date by $\pm1$ year, $\pm1$ month, or $\pm1$ day. \\
        Change Operator & \mutCO{} & Change \code{and} to \code{or} and vice versa. \\
        Negate Inequality & \mutNI{} & Change \code{=} to \code{!=} and vice versa. \\
        Reverse Inclusion & \mutRI{} & Change \code{in} condition to \code{notIn} or vice versa. \\
        Replace Startswith/Endswith & \mutRSE{} &  Change \code{startswith} to \code{endswith} and vice versa. \\
        Swap Rule & \mutSR{} & If \code{Rule A implies Rule B} then change it to \code{Rule B implies Rule A}. \\
        Swap Substring Indices & \mutSSI{} & If \code{substring(i,j)} then change it to \code{substring(j,i)}. \\
        \bottomrule
    \end{tabular}
\end{table}

%% file: rq1_metric.tex
Second, we study each \gls{llm}'s test generation success from two combined aspects:
\begin{inparaenum}
    \item how successful they are in producing exactly matched outputs
    and
    \item how many of the successfully matched outputs are true positives.
\end{inparaenum}
This is calculated as the Euclidean distance between the results of an \gls{llm} from these two aspects and the maximum possible values for them. Assuming the two points $A_{t}(\#Observed, \#True)$, and $B(T_{expected}, T_{expected})$. $A_{t}$ represents the results for an \gls{llm} for a rule for a test type $t$ (e.g., \rulePass{}), for $T_{expected}$ repetitions. $\#Observed$ is the number of times out of $T_{expected}$, the \gls{llm} produced exactly matched outputs, whereas $\#True$ represents how many matched outputs were true positives. Assuming $T_{expected}=30$, B will be $B(30,30)$. As an example---for point $A_{pass}$ (25, 15) means that we observe only 25 tests out of 30 (i.e., the maximum) for a specific rule under \rulePass{} and out of 25 observations only 15 tests are actually \textit{Passing} by \dvarepp{}.  
The Euclidean distance $ED_r$ is then defined in \cref{eq:edr}.
\begin{equation}
    \label{eq:edr}
    \begin{split}
        ED_r = \sqrt{(T_{expected} - \#Observed)^2 + (T_{expected} - \#True)^2}
    \end{split}
\end{equation}

$ED_r$ is then normalized using min-max normalization where the \textit{minimal distance} and \textit{maximal distance} are calculated based on the constant coordinates for distance calculation \((0, 0)\) and \((T_{expected}, T_{expected})\), respectively.
Finally, the success index $SI_r$ for a particular rule $r$ is calculated in \cref{eq:sir}.
\begin{equation}
\label{eq:sir}
SI_r = (1 - Normalized Distance) \times 100
\end{equation}

$SI_r$ represents the success percentage, with higher values indicating greater success.
Derived $SI_r$, we denote $SI_{pass}$, $SI_{fail}$, and $SI_{NotApplied}$ as the sets of success indices across all the rules per test type (\rulePass{}, \ruleFail{}, and \ruleNotApplied{}).

Third, we measure the time from sending the request to the \gls{llm} platform to receiving the response, including the model inference time and other communication overheads (e.g., network).
Throughout the paper, we refer to this time as the inference time \metricIT{}.

%% file: rq2_metric.tex
Let $RI_{rt}$ be the robustness index for a specific rule $r$ of test type $t$, as defined in \cref{eq:rit}.
\begin{equation}
    \label{eq:rit}
    \begin{split}
        RI_{rt} = 1 - \frac{1}{n_{rt}} \sum_{j=1}^{n_{rt}} \left| SI_{rt}^{original} - SI_{rtj}^{mutated} \right|
    \end{split}
\end{equation}
where
\begin{inparablank}
    \item $SI_{rt}^{\text{original}}$ is the success index for rule $r$ of test type $t$ on the original set;
    \item $SI_{rtj}^{\text{mutated}}$ is the success index for the $j$-th mutated rule for rule $r$ of test type $t$;
    and
    \item $n_{rt}$ is the total number of mutated rules for the original rule $r$ of test type $t$.
\end{inparablank}
Derived $RI_{rt}$, we denote $RI_{pass}$, $RI_{fail}$, and $RI_{NotApplied}$ as the sets of robustness indices across all the rules per test type (\rulePass{}, \ruleFail{}, and \ruleNotApplied{}).

%% file: rq3_metric.tex
Regarding differential testing, we use \textit{D} to denote the \dvarepp{} results and \textit{G} to denote the \guri{} results. We define the metrics for a match and a mismatch as follows:
$Match$ as the set of rule numbers $R$ associated with matches for a specific result, see \cref{eq:match_rules}.
A match is considered when both systems return the same result for the same test data.
\Cref{fig:approach} provides examples where, for a given rule and test, both system results should be the same (e.g., \rulePass, \ruleFail, or \ruleNotApplied).
\begin{equation}
\label{eq:match_rules}
    \begin{split}
     Match &= R \cap (D \cap G)
    \end{split}
\end{equation}
Similarly, we define $Mismatch$ in \cref{eq:mismatch_rules} as the set of rule numbers $R$ associated with mismatches for a specific result.
\begin{equation}
\label{eq:mismatch_rules}
    \begin{split}
     Mismatch &= R \cap (D \cup G)
    \end{split}
\end{equation}




%% file: rq1_table_stats.tex
\begin{table}[tbp]
    \ra{0.9}
    \centering
    \scriptsize
    \caption{Statistical tests of the completion rates, success indices per test type, and inference times per \gls{llm}}
    \label{tab:rq1:stats}
    \resizebox{\columnwidth}{!}{%
    \input{rq1_table_stats_body.tex}
    }
\end{table}

%% file: rq1_table_stats_body.tex
\begin{tabular}{llllrrl}
\toprule
Metric & Model 1 & Model 2 & Comp. & \pvalue{} & \vda{} & Magnitude \\
\midrule

\metricCR{} & \mistral{} & \llama{} & \statWorse{} & \pval{7.99e-88} & \vdaval{0.0586} & large \\
 & \mistral{} & \mixtral{} & \statEqual{} & \pval{0.653} & \vdaval{0.456} & negligible \\
 & \mistral{} & \gpt{} & \statWorse{} & \pval{6.39e-144} & \vdaval{0.0213} & large \\
 & \llama{} & \mixtral{} & \statBetter{} & \pval{1.44e-78} & \vdaval{0.939} & large \\
 & \llama{} & \gpt{} & \statWorse{} & \pval{3.46e-08} & \vdaval{0.303} & medium \\
 & \mixtral{} & \gpt{} & \statWorse{} & \pval{4.26e-132} & \vdaval{0.0215} & large \\
\midrule
\metricSIPass{} & \mistral{} & \llama{} & \statWorse{} & \pval{7.03e-33} & \vdaval{0.266} & medium \\
 & \mistral{} & \mixtral{} & \statWorse{} & \pval{0.000464} & \vdaval{0.369} & small \\
 & \mistral{} & \gpt{} & \statWorse{} & \pval{6.11e-72} & \vdaval{0.156} & large \\
 & \llama{} & \mixtral{} & \statBetter{} & \pval{2.69e-16} & \vdaval{0.699} & medium \\
 & \llama{} & \gpt{} & \statWorse{} & \pval{5.44e-09} & \vdaval{0.369} & small \\
 & \mixtral{} & \gpt{} & \statWorse{} & \pval{7.56e-46} & \vdaval{0.171} & large \\
\midrule
\metricSIFail{} & \mistral{} & \llama{} & \statWorse{} & \pval{5.85e-08} & \vdaval{0.404} & small \\
 & \mistral{} & \mixtral{} & \statWorse{} & \pval{1.43e-11} & \vdaval{0.312} & medium \\
 & \mistral{} & \gpt{} & \statWorse{} & \pval{3.23e-41} & \vdaval{0.234} & large \\
 & \llama{} & \mixtral{} & \statEqual{} & \pval{0.419} & \vdaval{0.488} & negligible \\
 & \llama{} & \gpt{} & \statWorse{} & \pval{7.11e-15} & \vdaval{0.337} & small \\
 & \mixtral{} & \gpt{} & \statWorse{} & \pval{9.92e-11} & \vdaval{0.335} & small \\
\midrule
\metricSINA{} & \mistral{} & \llama{} & \statWorse{} & \pval{1.03e-19} & \vdaval{0.314} & medium \\
 & \mistral{} & \mixtral{} & \statWorse{} & \pval{1.28e-05} & \vdaval{0.352} & small \\
 & \mistral{} & \gpt{} & \statWorse{} & \pval{8.34e-74} & \vdaval{0.143} & large \\
 & \llama{} & \mixtral{} & \statBetter{} & \pval{7.59e-06} & \vdaval{0.625} & small \\
 & \llama{} & \gpt{} & \statWorse{} & \pval{4.63e-19} & \vdaval{0.294} & medium \\
 & \mixtral{} & \gpt{} & \statWorse{} & \pval{3.02e-42} & \vdaval{0.177} & large \\
\midrule
\metricIT{} & \mistral{} & \llama{} & \statBetter{} & \pval{0} & \vdaval{0.234} & large \\
 & \mistral{} & \mixtral{} & \statEqual{} & \pval{3.32e-19} & \vdaval{0.544} & negligible \\
 & \mistral{} & \gpt{} & \statBetter{} & \pval{0} & \vdaval{0.173} & large \\
 & \llama{} & \mixtral{} & \statWorse{} & \pval{0} & \vdaval{0.82} & large \\
 & \llama{} & \gpt{} & \statBetter{} & \pval{3.63e-182} & \vdaval{0.346} & small \\
 & \mixtral{} & \gpt{} & \statBetter{} & \pval{0} & \vdaval{0.139} & large \\

\bottomrule
\end{tabular}

%% file: rq2_table_stats_rules.tex
\begin{table}[tbp]
    \ra{0.9}
    \centering
    \scriptsize
    \caption{Statistical tests of the robustness indices for all the rules per test type and \gls{llm}}
    \label{tab:rq2:stats:rules}
    \resizebox{\columnwidth}{!}{%
    \input{rq2_table_stats_rules_body.tex}
    }
\end{table}

%% file: rq2_table_stats_rules_body.tex
\begin{tabular}{llllrrl}
\toprule
Metric & Model 1 & Model 2 & Comp. & \pvalue{} & \vda{} & Magnitude \\
\midrule

\metricRIPass{} & \mistral{} & \llama{} & \statEqual{} & \pval{0.894} & \vdaval{0.457} & negligible \\
 & \mistral{} & \mixtral{} & \statEqual{} & \pval{0.894} & \vdaval{0.44} & negligible \\
 & \mistral{} & \gpt{} & \statWorse{} & \pval{3.36e-05} & \vdaval{0.236} & large \\
 & \llama{} & \mixtral{} & \statEqual{} & \pval{1} & \vdaval{0.486} & negligible \\
 & \llama{} & \gpt{} & \statWorse{} & \pval{0.00128} & \vdaval{0.325} & medium \\
 & \mixtral{} & \gpt{} & \statWorse{} & \pval{0.00128} & \vdaval{0.29} & medium \\
\midrule
\metricRIFail{} & \mistral{} & \llama{} & \statBetter{} & \pval{2.32e-06} & \vdaval{0.771} & large \\
 & \mistral{} & \mixtral{} & \statEqual{} & \pval{1} & \vdaval{0.526} & negligible \\
 & \mistral{} & \gpt{} & \statBetter{} & \pval{0.000558} & \vdaval{0.705} & medium \\
 & \llama{} & \mixtral{} & \statWorse{} & \pval{4.12e-06} & \vdaval{0.231} & large \\
 & \llama{} & \gpt{} & \statEqual{} & \pval{0.489} & \vdaval{0.414} & small \\
 & \mixtral{} & \gpt{} & \statBetter{} & \pval{0.00108} & \vdaval{0.708} & medium \\
\midrule
\metricRINA{} & \mistral{} & \llama{} & \statBetter{} & \pval{8.63e-05} & \vdaval{0.787} & large \\
 & \mistral{} & \mixtral{} & \statEqual{} & \pval{0.493} & \vdaval{0.39} & small \\
 & \mistral{} & \gpt{} & \statEqual{} & \pval{1} & \vdaval{0.444} & negligible \\
 & \llama{} & \mixtral{} & \statWorse{} & \pval{1.03e-07} & \vdaval{0.166} & large \\
 & \llama{} & \gpt{} & \statWorse{} & \pval{3.25e-06} & \vdaval{0.227} & large \\
 & \mixtral{} & \gpt{} & \statEqual{} & \pval{1} & \vdaval{0.531} & negligible \\

\bottomrule
\end{tabular}

%% file: rq3_table.tex
\begin{table*}[tbp]
    \ra{0.8}
    \centering
    \footnotesize
    \caption{Matches and mismatches between \guri{} and \dvarepp{} results, considering all 58 rules}
    \label{tab:rq3_results}
    \begin{tabular}{llrrl}
    \toprule
        \textbf{Model} & \textbf{\guri{} Result} & \multicolumn{2}{c}{\textbf{\dvarepp{} Result}} & \textbf{Mismatched Rules} \\ 
        & & \textbf{Match} & \textbf{Mismatch} & \\
        \midrule
        & \textit{Pass} & 44 (\perc{76}) & 14 (\perc{24.137931}) & V19/1, V19/2, V20/1, V20/2, V21, V25, V26, V27/1, V28, V29/1, V46, V50, V51, V67 \\
        & \textit{Fail} & 47 (\perc{81}) & 11 (\perc{18.965517}) & V17, V22/1, V22/2, V23, V24, V25, V27/1, V29/1, V46, V50, V51 \\
        \mistral & \textit{Not Applied} & 53 (\perc{91}) & 5 (\perc{8.62068966}) & V43, V44, V45, V46, V69/1 \\
        & \textit{Warning} & 57 (\perc{98}) & 1 (\perc{1.72413793}) & V08 \\
        & \textit{500} & 57 (\perc{98}) & 1 (\perc{1.72413793}) & V53 \\
        & \textit{Empty Response} & 46 (\perc{79}) & 12 (\perc{20.6896552}) & V17, V23, V24, V25, V26,  V28, V47,  V49, V51,  V53,  V67, V65   \\
        \midrule
        & \textit{Pass} & 52 (\perc{90}) & 6 (\perc{10.3448276}) & V19/2, V20/1, V20/2, V27/1, V46, V50,  \\
        & \textit{Fail} & 52 (\perc{90}) & 6 (\perc{10.3448276}) &  V19/1, V20/1, V22/1, V22/2, V27/1, V50 \\
        \llama & \textit{Not Applied} & 54 (\perc{93})& 4 (\perc{6.89655172}) &   V43, V44, V45, V69/1, \\
        & \textit{Warning} & 57 (\perc{98}) & 1 (\perc{1.72413793}) & V08 \\
        & \textit{500} & 57 (\perc{98}) & 1 (\perc{1.72413793}) & V53 \\
        & \textit{Empty Response} & 47 (\perc{81}) & 11 (\perc{18.965517}) & V17, V23, V24, V25, V26, V28, V47, V49, V51, V65, V67\\
        \midrule
        & \textit{Pass} & 55 (\perc{95}) & 3 (\perc{5.17241379})  & V22/2, V27/1, V46 \\
        & \textit{Fail} & 53 (\perc{91}) & 5 (\perc{8.62068966}) & V19/1, V22/1, V22/2, V27/1, V46 \\
        \mixtral & \textit{Not Applied} & 54 (\perc{93}) & 4 (\perc{6.89655172})  & V43, V44, V45, V69/1 \\
        & \textit{Warning} & 57 (\perc{98}) & 1 (\perc{1.72413793}) & V08 \\
        & \textit{500} & 56 (\perc{97}) & 2 (\perc{3.44827586}) & V47, V53 \\
        & \textit{Empty Response} & 46 (\perc{79})  & 12 (\perc{20.6896552}) &  V17, V23, V24, V25, V26, V28, V47, V49, V51, V53, V65, V67\\
        \midrule
        & \textit{Pass} & 52 (\perc{90}) & 6 (\perc{10.3448276}) & V20/2, V46, V50, V51, V64, V67 \\
        & \textit{Fail} & 49 (\perc{84})  & 9 (\perc{15.5172414}) & V19/1, V20/1, V22/1, V22/2, V24,  V26, V27/1, V46, V50 \\
        \gpt & \textit{Not Applied} & 53 (\perc{91}) & 5 (\perc{8.62068966}) & V43, V44, V45, V46, V69/1 \\
        & \textit{Warning} & 57 (\perc{98}) & 1 (\perc{1.72413793}) & V08 \\
        & \textit{500} & 57 (\perc{98}) & 1 (\perc{1.72413793}) & V53 \\
        & \textit{Empty Response} & 48 (\perc{83}) & 10 (\perc{17.2413793}) &  V17, V24, V25, V26, V28, V47, V51, V53, V65, V67\\
        \bottomrule
    \end{tabular}%
\end{table*}





%% file: main.bbl
\begin{thebibliography}{50}
\providecommand{\natexlab}[1]{#1}
\providecommand{\url}[1]{#1}
\csname url@samestyle\endcsname
\providecommand{\newblock}{\relax}
\providecommand{\bibinfo}[2]{#2}
\providecommand{\BIBentrySTDinterwordspacing}{\spaceskip=0pt\relax}
\providecommand{\BIBentryALTinterwordstretchfactor}{4}
\providecommand{\BIBentryALTinterwordspacing}{\spaceskip=\fontdimen2\font plus
\BIBentryALTinterwordstretchfactor\fontdimen3\font minus \fontdimen4\font\relax}
\providecommand{\BIBforeignlanguage}[2]{{%
\expandafter\ifx\csname l@#1\endcsname\relax
\typeout{** WARNING: IEEEtranSN.bst: No hyphenation pattern has been}%
\typeout{** loaded for the language `#1'. Using the pattern for}%
\typeout{** the default language instead.}%
\else
\language=\csname l@#1\endcsname
\fi
#2}}
\providecommand{\BIBdecl}{\relax}
\BIBdecl

\bibitem[Al{-}Kaswan et~al.(2023)Al{-}Kaswan, Ahmed, Izadi, Sawant, Devanbu, and van Deursen]{em:2}
\BIBentryALTinterwordspacing
A.~Al{-}Kaswan, T.~Ahmed, M.~Izadi, A.~A. Sawant, P.~T. Devanbu, and A.~van Deursen, ``Extending source code pre-trained language models to summarise decompiled binarie,'' in \emph{{IEEE} International Conference on Software Analysis, Evolution and Reengineering, {SANER} 2023, Taipa, Macao, March 21-24, 2023}.\hskip 1em plus 0.5em minus 0.4em\relax {IEEE}, 2023, pp. 260--271. [Online]. Available: \url{https://doi.org/10.1109/SANER56733.2023.00033}
\BIBentrySTDinterwordspacing

\bibitem[Arcuri(2019)]{arcuri:19}
\BIBentryALTinterwordspacing
A.~Arcuri, ``{REST}ful {API} automated test case generation with {E}vo{M}aster,'' \emph{{ACM} Transactions on Software Engineering and Methodology}, vol.~28, no.~1, pp. 1--37, Feb. 2019. [Online]. Available: \url{https://doi.org/10.1145/3293455}
\BIBentrySTDinterwordspacing

\bibitem[Arcuri and Briand(2011)]{arcuri:11}
\BIBentryALTinterwordspacing
A.~Arcuri and L.~Briand, ``A practical guide for using statistical tests to assess randomized algorithms in software engineering,'' in \emph{Proceedings of the 33rd International Conference on Software Engineering}, ser. {ICSE} 2011.\hskip 1em plus 0.5em minus 0.4em\relax {ACM}, 2011. [Online]. Available: \url{https://doi.org/10.1145/1985793.1985795}
\BIBentrySTDinterwordspacing

\bibitem[Barr et~al.(2015)Barr, Harman, McMinn, Shahbaz, and Yoo]{barr:15b}
\BIBentryALTinterwordspacing
E.~T. Barr, M.~Harman, P.~McMinn, M.~Shahbaz, and S.~Yoo, ``The oracle problem in software testing: A survey,'' \emph{{IEEE} Transactions on Software Engineering}, vol.~41, no.~5, pp. 507--525, May 2015. [Online]. Available: \url{https://doi.org/10.1109/tse.2014.2372785}
\BIBentrySTDinterwordspacing

\bibitem[Benjamini and Yekutieli(2001)]{benjamini:01}
\BIBentryALTinterwordspacing
Y.~Benjamini and D.~Yekutieli, ``The control of the false discovery rate in multiple testing under dependency,'' \emph{The Annals of Statistics}, vol.~29, no.~4, pp. 1165--1188, Aug. 2001. [Online]. Available: \url{https://doi.org/10.1214/aos/1013699998}
\BIBentrySTDinterwordspacing

\bibitem[Chakraborty et~al.(2022)Chakraborty, Ahmed, Ding, Devanbu, and Ray]{em:1}
\BIBentryALTinterwordspacing
S.~Chakraborty, T.~Ahmed, Y.~Ding, P.~T. Devanbu, and B.~Ray, ``Natgen: generative pre-training by "naturalizing" source code,'' in \emph{Proceedings of the 30th {ACM} Joint European Software Engineering Conference and Symposium on the Foundations of Software Engineering, {ESEC/FSE} 2022, Singapore, Singapore, November 14-18, 2022}.\hskip 1em plus 0.5em minus 0.4em\relax {ACM}, 2022, pp. 18--30. [Online]. Available: \url{https://doi.org/10.1145/3540250.3549162}
\BIBentrySTDinterwordspacing

\bibitem[Chen et~al.(2018)Chen, Kuo, Liu, Poon, Towey, Tse, and Zhou]{tsong:18}
\BIBentryALTinterwordspacing
T.~Y. Chen, F.~Kuo, H.~Liu, P.~Poon, D.~Towey, T.~H. Tse, and Z.~Q. Zhou, ``Metamorphic testing: {A} review of challenges and opportunities,'' \emph{{ACM} Computing Surveys}, vol.~51, no.~1, pp. 4:1--4:27, 2018. [Online]. Available: \url{https://doi.org/10.1145/3143561}
\BIBentrySTDinterwordspacing

\bibitem[Chung et~al.(2023)Chung, Kamar, and Amershi]{chung:23}
\BIBentryALTinterwordspacing
J.~J.~Y. Chung, E.~Kamar, and S.~Amershi, ``Increasing diversity while maintaining accuracy: Text data generation with large language models and human interventions,'' in \emph{Proceedings of the 61st Annual Meeting of the Association for Computational Linguistics (Volume 1: Long Papers), {ACL} 2023, Toronto, Canada, July 9-14, 2023}.\hskip 1em plus 0.5em minus 0.4em\relax Association for Computational Linguistics, 2023, pp. 575--593. [Online]. Available: \url{https://doi.org/10.18653/v1/2023.acl-long.34}
\BIBentrySTDinterwordspacing

\bibitem[Clusmann et~al.(2023)Clusmann, Kolbinger, Muti, Carrero, Eckardt, Laleh, L{\"o}ffler, Schwarzkopf, Unger, Veldhuizen, et~al.]{clusmann:23}
J.~Clusmann, F.~R. Kolbinger, H.~S. Muti, Z.~I. Carrero, J.-N. Eckardt, N.~G. Laleh, C.~M.~L. L{\"o}ffler, S.-C. Schwarzkopf, M.~Unger, G.~P. Veldhuizen \emph{et~al.}, ``The future landscape of large language models in medicine,'' \emph{Communications Medicine}, vol.~3, no.~1, p. 141, 2023.

\bibitem[Dakhel et~al.(2023)Dakhel, Nikanjam, Majdinasab, Khomh, and Desmarais]{moradi:23}
\BIBentryALTinterwordspacing
A.~M. Dakhel, A.~Nikanjam, V.~Majdinasab, F.~Khomh, and M.~C. Desmarais, ``Effective test generation using pre-trained large language models and mutation testing,'' \emph{CoRR}, vol. abs/2308.16557, 2023. [Online]. Available: \url{https://doi.org/10.48550/arXiv.2308.16557}
\BIBentrySTDinterwordspacing

\bibitem[Deng et~al.(2023)Deng, Xia, Peng, Yang, and Zhang]{deng:23}
\BIBentryALTinterwordspacing
Y.~Deng, C.~S. Xia, H.~Peng, C.~Yang, and L.~Zhang, ``Large language models are zero-shot fuzzers: Fuzzing deep-learning libraries via large language models,'' in \emph{Proceedings of the 32nd {ACM} {SIGSOFT} International Symposium on Software Testing and Analysis, {ISSTA} 2023}.\hskip 1em plus 0.5em minus 0.4em\relax {ACM}, 2023, pp. 423--435. [Online]. Available: \url{https://doi.org/10.1145/3597926.3598067}
\BIBentrySTDinterwordspacing

\bibitem[Dunn(1964)]{dunn:64}
\BIBentryALTinterwordspacing
O.~J. Dunn, ``Multiple comparisons using rank sums,'' \emph{Technometrics}, vol.~6, no.~3, pp. 241--252, Aug. 1964. [Online]. Available: \url{https://doi.org/10.1080/00401706.1964.10490181}
\BIBentrySTDinterwordspacing

\bibitem[et~al.(2023{\natexlab{a}})]{mistral:23}
\BIBentryALTinterwordspacing
A.~Q.~J. et~al., ``Mistral 7b,'' \emph{CoRR}, vol. abs/2310.06825, 2023. [Online]. Available: \url{https://doi.org/10.48550/arXiv.2310.06825}
\BIBentrySTDinterwordspacing

\bibitem[et~al.(2024)]{mixtral:24}
\BIBentryALTinterwordspacing
------, ``Mixtral of experts,'' \emph{CoRR}, vol. abs/2401.04088, 2024. [Online]. Available: \url{https://doi.org/10.48550/arXiv.2401.04088}
\BIBentrySTDinterwordspacing

\bibitem[et~al.(2023{\natexlab{c}})]{peng:23}
\BIBentryALTinterwordspacing
C.~P. et~al., ``A study of generative large language model for medical research and healthcare,'' \emph{npj Digitital Medicine}, vol.~6, 2023. [Online]. Available: \url{https://doi.org/10.1038/s41746-023-00958-w}
\BIBentrySTDinterwordspacing

\bibitem[et~al.(2023{\natexlab{b}})]{llama:23}
\BIBentryALTinterwordspacing
H.~T. et~al., ``Llama 2: Open foundation and fine-tuned chat models,'' \emph{CoRR}, vol. abs/2307.09288, 2023. [Online]. Available: \url{https://doi.org/10.48550/arXiv.2307.09288}
\BIBentrySTDinterwordspacing

\bibitem[Gao et~al.(2023)Gao, Wen, Gao, Wang, Zhang, and Lyu]{em:3}
\BIBentryALTinterwordspacing
S.~Gao, X.~Wen, C.~Gao, W.~Wang, H.~Zhang, and M.~R. Lyu, ``What makes good in-context demonstrations for code intelligence tasks with llms?'' in \emph{38th {IEEE/ACM} International Conference on Automated Software Engineering, {ASE} 2023, Luxembourg, September 11-15, 2023}.\hskip 1em plus 0.5em minus 0.4em\relax {IEEE}, 2023, pp. 761--773. [Online]. Available: \url{https://doi.org/10.1109/ASE56229.2023.00109}
\BIBentrySTDinterwordspacing

\bibitem[Godefroid et~al.(2020)Godefroid, Lehmann, and Polishchuk]{godefroid:20}
\BIBentryALTinterwordspacing
P.~Godefroid, D.~Lehmann, and M.~Polishchuk, ``Differential regression testing for {REST} apis,'' in \emph{{ISSTA} '20: 29th {ACM} {SIGSOFT} International Symposium on Software Testing and Analysis, Virtual Event, USA, July 18-22, 2020}.\hskip 1em plus 0.5em minus 0.4em\relax {ACM}, 2020, pp. 312--323. [Online]. Available: \url{https://doi.org/10.1145/3395363.3397374}
\BIBentrySTDinterwordspacing

\bibitem[Guilherme and Vincenzi(2023)]{vitor:23}
\BIBentryALTinterwordspacing
V.~Guilherme and A.~Vincenzi, ``An initial investigation of chatgpt unit test generation capability,'' in \emph{8th Brazilian Symposium on Systematic and Automated Software Testing, {SAST} 2023, Campo Grande, MS, Brazil, September 25-29, 2023}.\hskip 1em plus 0.5em minus 0.4em\relax {ACM}, 2023, pp. 15--24. [Online]. Available: \url{https://doi.org/10.1145/3624032.3624035}
\BIBentrySTDinterwordspacing

\bibitem[Hamlet(2021)]{hamlet:21}
\BIBentryALTinterwordspacing
R.~G. Hamlet, ``Random testing,'' \emph{Essentials of Software Testing}, 2021. [Online]. Available: \url{https://api.semanticscholar.org/CorpusID:6665543}
\BIBentrySTDinterwordspacing

\bibitem[Hess and Kromrey(2004)]{hess:04}
M.~R. Hess and J.~D. Kromrey, ``Robust confidence intervals for effect sizes: A comparative study of cohen's d and cliff's delta under non-normality and heterogeneous variances,'' \emph{Annual Meeting of the American Educational Research Association}, Apr. 2004.

\bibitem[Huang et~al.(2023)Huang, Yu, Ma, Zhong, Feng, Wang, Chen, Peng, Feng, Qin, and Liu]{huang:23}
\BIBentryALTinterwordspacing
L.~Huang, W.~Yu, W.~Ma, W.~Zhong, Z.~Feng, H.~Wang, Q.~Chen, W.~Peng, X.~Feng, B.~Qin, and T.~Liu, ``A survey on hallucination in large language models: Principles, taxonomy, challenges, and open questions,'' \emph{CoRR}, vol. abs/2311.05232, 2023. [Online]. Available: \url{https://doi.org/10.48550/arXiv.2311.05232}
\BIBentrySTDinterwordspacing

\bibitem[Hyun et~al.(2023)Hyun, Guo, and Babar]{hyun:23}
\BIBentryALTinterwordspacing
S.~Hyun, M.~Guo, and M.~A. Babar, ``{METAL:} metamorphic testing framework for analyzing large-language model qualities,'' \emph{CoRR}, vol. abs/2312.06056, 2023. [Online]. Available: \url{https://doi.org/10.48550/arXiv.2312.06056}
\BIBentrySTDinterwordspacing

\bibitem[Isaku et~al.(2023)Isaku, Sartaj, Laaber, Ali, Yue, Schwitalla, and Nyg{\aa}rd]{isaku:23}
E.~Isaku, H.~Sartaj, C.~Laaber, S.~Ali, T.~Yue, T.~Schwitalla, and J.~F. Nyg{\aa}rd, ``Cost reduction on testing evolving cancer registry system,'' in \emph{Proceedings of the 39th {IEEE} International Conference on Software Maintenance and Evolution}, ser. {ICSME} 2023.\hskip 1em plus 0.5em minus 0.4em\relax {IEEE}, Oct. 2023.

\bibitem[Jia and Harman(2011)]{jia:11}
\BIBentryALTinterwordspacing
Y.~Jia and M.~Harman, ``An analysis and survey of the development of mutation testing,'' \emph{{IEEE} Transactions on Software Engineering}, vol.~37, no.~5, pp. 649--678, Sep. 2011. [Online]. Available: \url{http://dx.doi.org/10.1109/TSE.2010.62}
\BIBentrySTDinterwordspacing

\bibitem[Kruskal and Wallis(1952)]{kruskal:52}
\BIBentryALTinterwordspacing
W.~H. Kruskal and W.~A. Wallis, ``Use of ranks in one-criterion variance analysis,'' \emph{Journal of the American Statistical Association}, vol.~47, no. 260, pp. 583--621, Dec. 1952. [Online]. Available: \url{https://doi.org/10.1080/01621459.1952.10483441}
\BIBentrySTDinterwordspacing

\bibitem[Laaber et~al.(2023{\natexlab{a}})Laaber, Yue, Ali, Schwitalla, and Nyg{\aa}rd]{laaber:23a}
\BIBentryALTinterwordspacing
C.~Laaber, T.~Yue, S.~Ali, T.~Schwitalla, and J.~F. Nyg{\aa}rd, ``Automated test generation for medical rules web services: A case study at the {C}ancer {R}egistry of {N}orway,'' in \emph{Proceedings of the 31st {ACM} Joint European Software Engineering Conference and Symposium on the Foundations of Software Engineering}, ser. {ESEC}/{FSE} 2023.\hskip 1em plus 0.5em minus 0.4em\relax {ACM}, Dec. 2023. [Online]. Available: \url{https://doi.org/10.1145/3611643.3613882}
\BIBentrySTDinterwordspacing

\bibitem[Laaber et~al.(2023{\natexlab{b}})Laaber, Yue, Ali, Schwitalla, and Nyg{\aa}rd]{laaber:23b}
\BIBentryALTinterwordspacing
------, ``Challenges of testing an evolving cancer registration support system in practice,'' in \emph{Proceedings of the 45th {IEEE}/{ACM} International Conference on Software Engineering: Companion Proceedings}, ser. {ICSE}-Companion 2023.\hskip 1em plus 0.5em minus 0.4em\relax {IEEE}, May 2023, pp. 355--359. [Online]. Available: \url{https://doi.org/10.1109/ICSE-Companion58688.2023.00102}
\BIBentrySTDinterwordspacing

\bibitem[Leinonen et~al.(2023)Leinonen, Hellas, Sarsa, Reeves, Denny, Prather, and Becker]{juho:23}
\BIBentryALTinterwordspacing
J.~Leinonen, A.~Hellas, S.~Sarsa, B.~N. Reeves, P.~Denny, J.~Prather, and B.~A. Becker, ``Using large language models to enhance programming error messages,'' in \emph{Proceedings of the 54th {ACM} Technical Symposium on Computer Science Education, Volume 1, {SIGCSE} 2023, Toronto, ON, Canada, March 15-18, 2023}.\hskip 1em plus 0.5em minus 0.4em\relax {ACM}, 2023, pp. 563--569. [Online]. Available: \url{https://doi.org/10.1145/3545945.3569770}
\BIBentrySTDinterwordspacing

\bibitem[Lu et~al.(2023)Lu, Xu, Yue, Ali, Schwitalla, and Nyg{\aa}rd]{chengjie:23}
\BIBentryALTinterwordspacing
C.~Lu, Q.~Xu, T.~Yue, S.~Ali, T.~Schwitalla, and J.~Nyg{\aa}rd, ``Evoclinical: Evolving cyber-cyber digital twin with active transfer learning for automated cancer registry system,'' in \emph{Proceedings of the 31st {ACM} Joint European Software Engineering Conference and Symposium on the Foundations of Software Engineering, {ESEC/FSE} 2023, San Francisco, CA, USA, December 3-9, 2023}.\hskip 1em plus 0.5em minus 0.4em\relax {ACM}, 2023, pp. 1973--1984. [Online]. Available: \url{https://doi.org/10.1145/3611643.3613897}
\BIBentrySTDinterwordspacing

\bibitem[Lu et~al.(2019)Lu, Wang, Yue, Ali, and Nyg{\aa}rd]{lu:19}
\BIBentryALTinterwordspacing
H.~Lu, S.~Wang, T.~Yue, S.~Ali, and J.~F. Nyg{\aa}rd, ``Automated refactoring of {OCL} constraints with search,'' \emph{{IEEE} Trans. Software Eng.}, vol.~45, no.~2, pp. 148--170, 2019. [Online]. Available: \url{https://doi.org/10.1109/TSE.2017.2774829}
\BIBentrySTDinterwordspacing

\bibitem[McKeeman(1998)]{mckeeman:98}
\BIBentryALTinterwordspacing
W.~M. McKeeman, ``Differential testing for software,'' \emph{Digital Technical Journal}, vol.~10, no.~1, pp. 100--107, 1998. [Online]. Available: \url{https://www.hpl.hp.com/hpjournal/dtj/vol10num1/vol10num1art9.pdf}
\BIBentrySTDinterwordspacing

\bibitem[{O}pen{AI}(2024)]{gpt}
\BIBentryALTinterwordspacing
{O}pen{AI}, ``{GPT} 3.5,'' 2024. [Online]. Available: \url{https://platform.openai.com/docs/models/gpt-3-5}
\BIBentrySTDinterwordspacing

\bibitem[OpenAI(2023)]{gpt-4}
\BIBentryALTinterwordspacing
OpenAI, ``{GPT-4} technical report,'' \emph{CoRR}, vol. abs/2303.08774, 2023. [Online]. Available: \url{https://doi.org/10.48550/arXiv.2303.08774}
\BIBentrySTDinterwordspacing

\bibitem[Segura et~al.(2018)Segura, Parejo, Troya, and Cort{\'{e}}s]{segura:18}
\BIBentryALTinterwordspacing
S.~Segura, J.~A. Parejo, J.~Troya, and A.~R. Cort{\'{e}}s, ``Metamorphic testing of restful web apis,'' \emph{{IEEE} Trans. Software Eng.}, vol.~44, no.~11, pp. 1083--1099, 2018. [Online]. Available: \url{https://doi.org/10.1109/TSE.2017.2764464}
\BIBentrySTDinterwordspacing

\bibitem[Shan and Zhu(2009)]{shan:09}
\BIBentryALTinterwordspacing
L.~Shan and H.~Zhu, ``Generating structurally complex test cases by data mutation: A case study of testing an automated modelling tool,'' \emph{The Computer Journal}, vol.~52, no.~5, pp. 571--588, 2009. [Online]. Available: \url{https://doi.org/10.1093/comjnl/bxm043}
\BIBentrySTDinterwordspacing

\bibitem[Stol and Fitzgerald(2018)]{stol:18}
\BIBentryALTinterwordspacing
K.-J. Stol and B.~Fitzgerald, ``The {ABC} of software engineering research,'' \emph{{ACM} Transactions on Software Engineering and Methodology}, vol.~27, no.~3, pp. 1--51, Oct. 2018. [Online]. Available: \url{https://doi.org/10.1145/3241743}
\BIBentrySTDinterwordspacing

\bibitem[Thirunavukarasu et~al.(2023)Thirunavukarasu, Ting, Elangovan, Gutierrez, Tan, and Ting]{thirunavukarasu:23}
A.~J. Thirunavukarasu, D.~S.~J. Ting, K.~Elangovan, L.~Gutierrez, T.~F. Tan, and D.~S.~W. Ting, ``Large language models in medicine,'' \emph{Nature medicine}, vol.~29, no.~8, pp. 1930--1940, 2023.

\bibitem[Tufano et~al.(2022)Tufano, Drain, Svyatkovskiy, and Sundaresan]{tufano:22}
\BIBentryALTinterwordspacing
M.~Tufano, D.~Drain, A.~Svyatkovskiy, and N.~Sundaresan, ``Generating accurate assert statements for unit test cases using pretrained transformers,'' in \emph{{IEEE/ACM} International Conference on Automation of Software Test, AST@ICSE 2022, Pittsburgh, PA, USA, May 21-22, 2022}.\hskip 1em plus 0.5em minus 0.4em\relax {ACM/IEEE}, 2022, pp. 54--64. [Online]. Available: \url{https://doi.org/10.1145/3524481.3527220}
\BIBentrySTDinterwordspacing

\bibitem[Vargha and Delaney(2000)]{vargha:00}
\BIBentryALTinterwordspacing
A.~Vargha and H.~D. Delaney, ``A critique and improvement of the "{CL}" common language effect size statistics of {McGraw} and {Wong},'' \emph{Journal of Educational and Behavioral Statistics}, vol.~25, no.~2, pp. 101--132, 2000. [Online]. Available: \url{https://doi.org/10.2307/1165329}
\BIBentrySTDinterwordspacing

\bibitem[Vikram et~al.(2023)Vikram, Lemieux, and Padhye]{vasudev:23}
\BIBentryALTinterwordspacing
V.~Vikram, C.~Lemieux, and R.~Padhye, ``Can large language models write good property-based tests?'' \emph{CoRR}, vol. abs/2307.04346, 2023. [Online]. Available: \url{https://doi.org/10.48550/arXiv.2307.04346}
\BIBentrySTDinterwordspacing

\bibitem[Wang et~al.(2023)Wang, Huang, Chen, Liu, Wang, and Wang]{wang:23}
\BIBentryALTinterwordspacing
J.~Wang, Y.~Huang, C.~Chen, Z.~Liu, S.~Wang, and Q.~Wang, ``Software testing with large language model: Survey, landscape, and vision,'' \emph{CoRR}, vol. abs/2307.07221, 2023. [Online]. Available: \url{https://doi.org/10.48550/arXiv.2307.07221}
\BIBentrySTDinterwordspacing

\bibitem[Wang et~al.(2016)Wang, Lu, Yue, Ali, and Nyg{\aa}rd]{wang:16}
\BIBentryALTinterwordspacing
S.~Wang, H.~Lu, T.~Yue, S.~Ali, and J.~Nyg{\aa}rd, ``{MBF4CR:} {A} model-based framework for supporting an automated cancer registry system,'' in \emph{Modelling Foundations and Applications - 12th European Conference, ECMFA@STAF 2016, Vienna, Austria, July 6-7, 2016, Proceedings}, ser. Lecture Notes in Computer Science, vol. 9764.\hskip 1em plus 0.5em minus 0.4em\relax Springer, 2016, pp. 191--204. [Online]. Available: \url{https://doi.org/10.1007/978-3-319-42061-5\_12}
\BIBentrySTDinterwordspacing

\bibitem[Wang et~al.(2017)Wang, Schwitalla, Yue, Ali, and Nyg{\aa}rd]{wang:17}
\BIBentryALTinterwordspacing
S.~Wang, T.~Schwitalla, T.~Yue, S.~Ali, and J.~F. Nyg{\aa}rd, ``{RCIA:} automated change impact analysis to facilitate a practical cancer registry system,'' in \emph{2017 {IEEE} International Conference on Software Maintenance and Evolution, {ICSME} 2017, Shanghai, China, September 17-22, 2017}.\hskip 1em plus 0.5em minus 0.4em\relax {IEEE} Computer Society, 2017, pp. 603--612. [Online]. Available: \url{https://doi.org/10.1109/ICSME.2017.22}
\BIBentrySTDinterwordspacing

\bibitem[Wornow et~al.(2023)Wornow, Xu, Thapa, Patel, Steinberg, Fleming, Pfeffer, Fries, and Shah]{xu:23}
\BIBentryALTinterwordspacing
M.~Wornow, Y.~Xu, R.~Thapa, B.~S. Patel, E.~Steinberg, S.~L. Fleming, M.~A. Pfeffer, J.~A. Fries, and N.~H. Shah, ``The shaky foundations of large language models and foundation models for electronic health records,'' \emph{npj Digitital Medicine}, vol.~6, 2023. [Online]. Available: \url{https://doi.org/10.1038/s41746-023-00879-8}
\BIBentrySTDinterwordspacing

\bibitem[Xie et~al.(2023)Xie, Chen, Zhi, Deng, and Yin]{xie:23}
\BIBentryALTinterwordspacing
Z.~Xie, Y.~Chen, C.~Zhi, S.~Deng, and J.~Yin, ``Chatunitest: a chatgpt-based automated unit test generation tool,'' \emph{CoRR}, vol. abs/2305.04764, 2023. [Online]. Available: \url{https://doi.org/10.48550/arXiv.2305.04764}
\BIBentrySTDinterwordspacing

\bibitem[Xu et~al.(2022)Xu, Alon, Neubig, and Hellendoorn]{temp:22}
\BIBentryALTinterwordspacing
F.~F. Xu, U.~Alon, G.~Neubig, and V.~J. Hellendoorn, ``A systematic evaluation of large language models of code,'' in \emph{Proceedings of the 6th {ACM} {SIGPLAN} International Symposium on Machine Programming}, ser. MAPS@PLDI 2022.\hskip 1em plus 0.5em minus 0.4em\relax {ACM}, 2022, pp. 1--10. [Online]. Available: \url{https://doi.org/10.1145/3520312.3534862}
\BIBentrySTDinterwordspacing

\bibitem[Zhang et~al.(2023{\natexlab{b}})Zhang, Vemulapalli, Talukdar, Ahn, Wang, Meng, Murtaza, Dave, Leshchiner, Joseph, Witteveen{-}Lane, Chesla, Zhou, and Chen]{zhang:23}
\BIBentryALTinterwordspacing
X.~Zhang, S.~Vemulapalli, N.~Talukdar, S.~Ahn, J.~Wang, H.~Meng, S.~M.~B. Murtaza, A.~A. Dave, D.~Leshchiner, D.~F. Joseph, M.~Witteveen{-}Lane, D.~Chesla, J.~Zhou, and B.~Chen, ``Large language models in medical term classification and unexpected misalignment between response and reasoning,'' \emph{CoRR}, vol. abs/2312.14184, 2023. [Online]. Available: \url{https://doi.org/10.48550/arXiv.2312.14184}
\BIBentrySTDinterwordspacing

\bibitem[Zhang et~al.(2023{\natexlab{a}})Zhang, Yang, Yuan, and Yao]{yifan:23}
\BIBentryALTinterwordspacing
Y.~Zhang, J.~Yang, Y.~Yuan, and A.~C. Yao, ``Cumulative reasoning with large language models,'' \emph{CoRR}, vol. abs/2308.04371, 2023. [Online]. Available: \url{https://doi.org/10.48550/arXiv.2308.04371}
\BIBentrySTDinterwordspacing

\bibitem[Zhu et~al.(1997)Zhu, Hall, and May]{hong:97}
\BIBentryALTinterwordspacing
H.~Zhu, P.~A.~V. Hall, and J.~H.~R. May, ``Software unit test coverage and adequacy,'' \emph{{ACM} Computing Surveys}, vol.~29, no.~4, pp. 366--427, 1997. [Online]. Available: \url{https://doi.org/10.1145/267580.267590}
\BIBentrySTDinterwordspacing

\end{thebibliography}
